\documentclass[prc, showpacs, showkeys, superscriptaddress, preprint]{revtex4}

\usepackage{graphicx}

\newcommand{\nuc}[2]{\ensuremath{^{\text{#1}}\text{#2}}}
\newcommand{\nucc}[3]{\ensuremath{^{\text{#1}}\text{#2}^{\text{#3}}}}
\newcommand{\mevc}{\mbox{MeV\hspace{-0.111em}/$c$}}
\newcommand{\mevu}{\mbox{MeV/nucleon}}

\newcommand{\nscl}{\affiliation{National Superconducting Cyclotron Laboratory, Michigan State University, East Lansing, Michigan 48824, USA}}
\newcommand{\pa}{\affiliation{Department of Physics \& Astronomy, Michigan State University, East Lansing, Michigan 48824, USA}}
\newcommand{\lan}{\affiliation{Institute of Modern Physics, CAS, Lanzhou 730000, China}}
\newcommand{\riken}{\affiliation{RIKEN Nishina Center, 2-1 Hirosawa, Wako, Saitama 351-0198 Japan}}
\newcommand{\uot}{\affiliation{Department of Physics, University of Tokyo, 7-3-1 Hongo, Bunkyo, Tokyo 113-0033, Japan}}
\newcommand{\cns}{\affiliation{Center for Nuclear Study, University of Tokyo (CNS), RIKEN campus, 2-1, Hirosawa, Wako, Saitama 351-0198, Japan}}
\newcommand{\rikkyo}{\affiliation{Rikkyo University, 3 Nishi-Ikebukuro, Toshima, Tokyo 171, Japan}}

\begin{document}
\title{Projectile Fragmentation of \nuc{86}{Kr} at 64 \mevu}
\author{M.~Mocko}
\email[Corresponding author: ]{mmocko@lanl.gov}\nscl\pa
\author{M.~B.~Tsang}\nscl\pa
\author{Z.~Y.~Sun}\lan
\author{N.~Aoi}\riken
\author{J.~Cook}\nscl\pa
\author{F.~Delaunay}\nscl
\author{M.~A.~Famiano}\nscl
\author{H.~Hui}\nscl
\author{N.~Imai}\riken
\author{H.~Iwasaki}\uot
\author{W.~G.~Lynch}\nscl\pa
\author{T.~Motobayashi}\riken
\author{M.~Niikura}\cns
\author{T.~Onishi}\uot
\author{A.~M.~Rogers}\nscl\pa
\author{H.~Sakurai}\uot
\author{A.~Stolz}\nscl
\author{H.~Suzuki}\uot
\author{E.~Takeshita}\rikkyo
\author{S.~Takeuchi}\riken
\author{M.~S.~Wallace}\nscl\pa

\date{\today}
\begin{abstract}
We measured fragmentation cross sections produced using the primary beam of \nuc{86}{Kr} at 64 \mevu\, on \nuc{9}{Be} and \nuc{181}{Ta} targets. The cross sections were obtained by integrating the momentum distributions of isotopes with $25\leq Z\leq 36$ measured using the RIPS fragment separator at RIKEN. The cross-section ratios obtained with the \nuc{181}{Ta} and \nuc{9}{Be} targets depend on the fragment masses, contrary to the simple geometrical models. We compared the extracted cross sections to EPAX; an empirical parameterization of fragmentation cross sections. Predictions from current EPAX parameterization severely overestimate the production cross sections of very neutron-rich isotopes.  Attempts to obtain another set of EPAX parameters specific to the reaction studied here, to extrapolate the neutron-rich nuclei more accurately have not been very successful, suggesting that accurate predictions of production cross sections of nuclei far from the valley of stability require information of nuclear properties which are not present in EPAX. 
\end{abstract}



\maketitle

\section{Introduction}
With recent developments in heavy-ion accelerators and rare isotope beam production many new surprising phenomena have been observed in unstable nuclei, such as neutron halo \cite{tan91}, neutron and proton skins of nuclei far from stability \cite{tan92, fuk93}, and large deformations of neutron-rich isotopes \cite{mot95}. In the planning and development of experiments with rare isotope beams, the EPAX code is used extensively in the current radioactive ion beam facilities. EPAX is an empirical parameterization of fragmentation cross sections relying on data mainly from reactions at incident energy greater than 200 \mevu. Using EPAX at low incident energy assumes the validity of limiting fragmentation, when the production cross sections do not depend on incident energy or target. It is, therefore, very important to verify EPAX predictions of production of rare isotopes at extreme proton and neutron compositions, especially for facilities that produce radioactive ion beams at incident energies lower than 200 \mevu. 

The present study compares fragment production cross sections from the projectile fragmentation of \nuc{86}{Kr} at 64 \mevu\, to EPAX, an empirical parameterization of fragmentation cross sections. \nuc{86}{Kr} is chosen as it is one of the most neutron-rich naturally occurring stable isotopes. Due to its noble gas chemical properties and that it can be easily ionized in an ion source, projectile fragmentation of \nuc{86}{Kr} is widely used to produce neutron-rich rare isotopes.  

\section{Experimental setup}
\begin{figure}
\includegraphics[width=\textwidth]{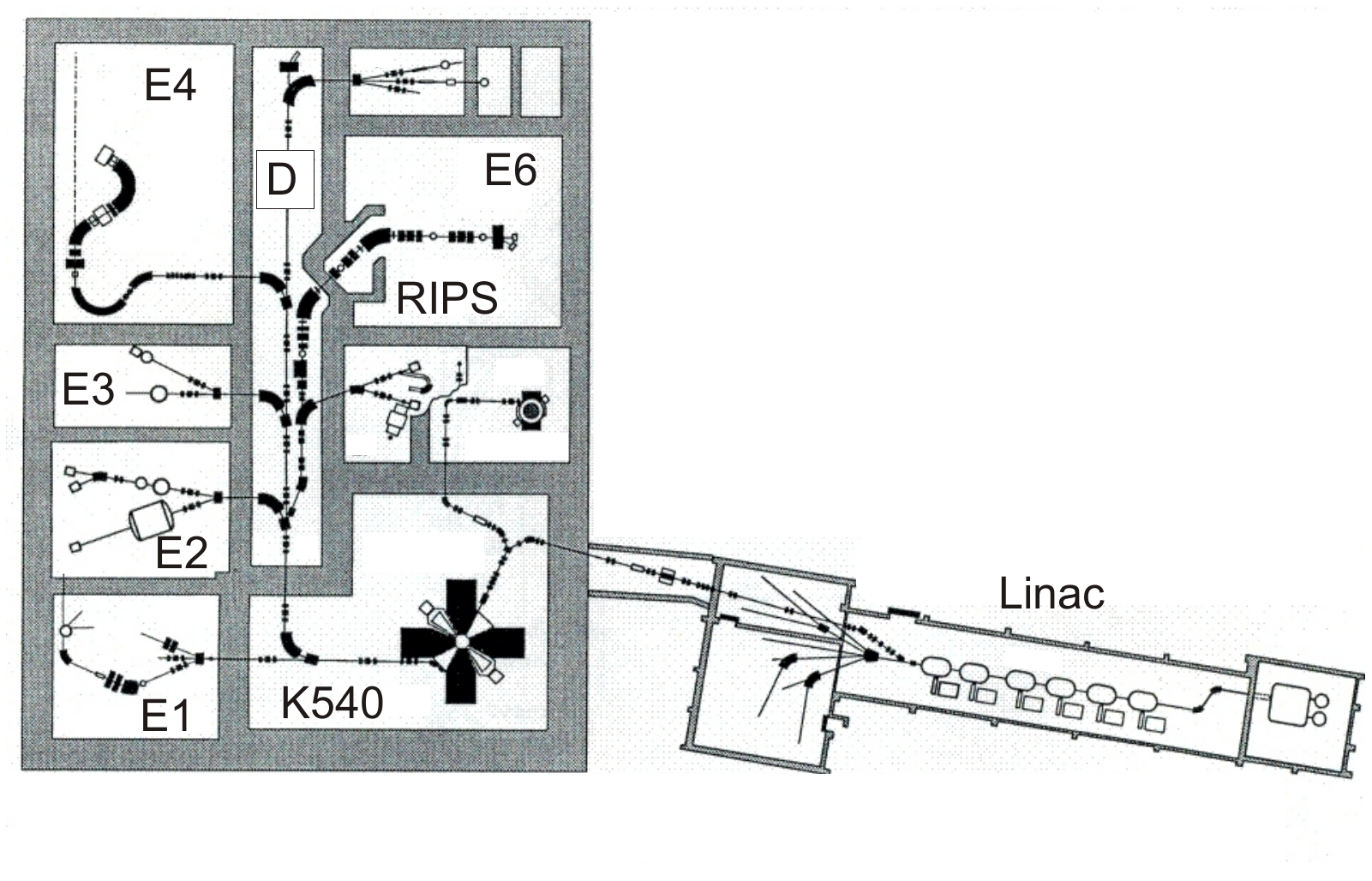}
\caption{Layout of the experimental facility at RIKEN. The LINAC injector and the K540 cyclotron are shown along with the experimental areas E1--E6. The RIPS fragment separator is located in experimental areas D and E6. \cite{kub92}}\label{fig1}
\end{figure}

The fragmentation experiments were carried out at RIKEN Accelerator Research Facility \cite{yan89}. A primary beam of \nuc{86}{Kr} with incident energy of 64 \mevu\, was produced by injecting \nuc{86}{Kr} ions into the K540 Ring Cyclotron using the LINAC injector. The layout of the LINAC, K540 Ring Cyclotron, and the experimental areas in the RIKEN facility is shown in Fig. \ref{fig1}. 
Two reaction targets, 96 mg/cm$^2$ \nuc{9}{Be} and 156 mg/cm$^2$ \nuc{181}{Ta} foils, were used. The target thicknesses were chosen such that the energy losses of the primary beam in the targets were similar thus data could be taken with both targets using the same magnetic setting. Minimizing the number of settings required in the experiments results in better utilization of the primary beam since changing the magnetic setting of the RIKEN Projectile Fragment Separator (RIPS) takes much longer than changing the targets.

\begin{figure}
\includegraphics[width=0.5\textwidth]{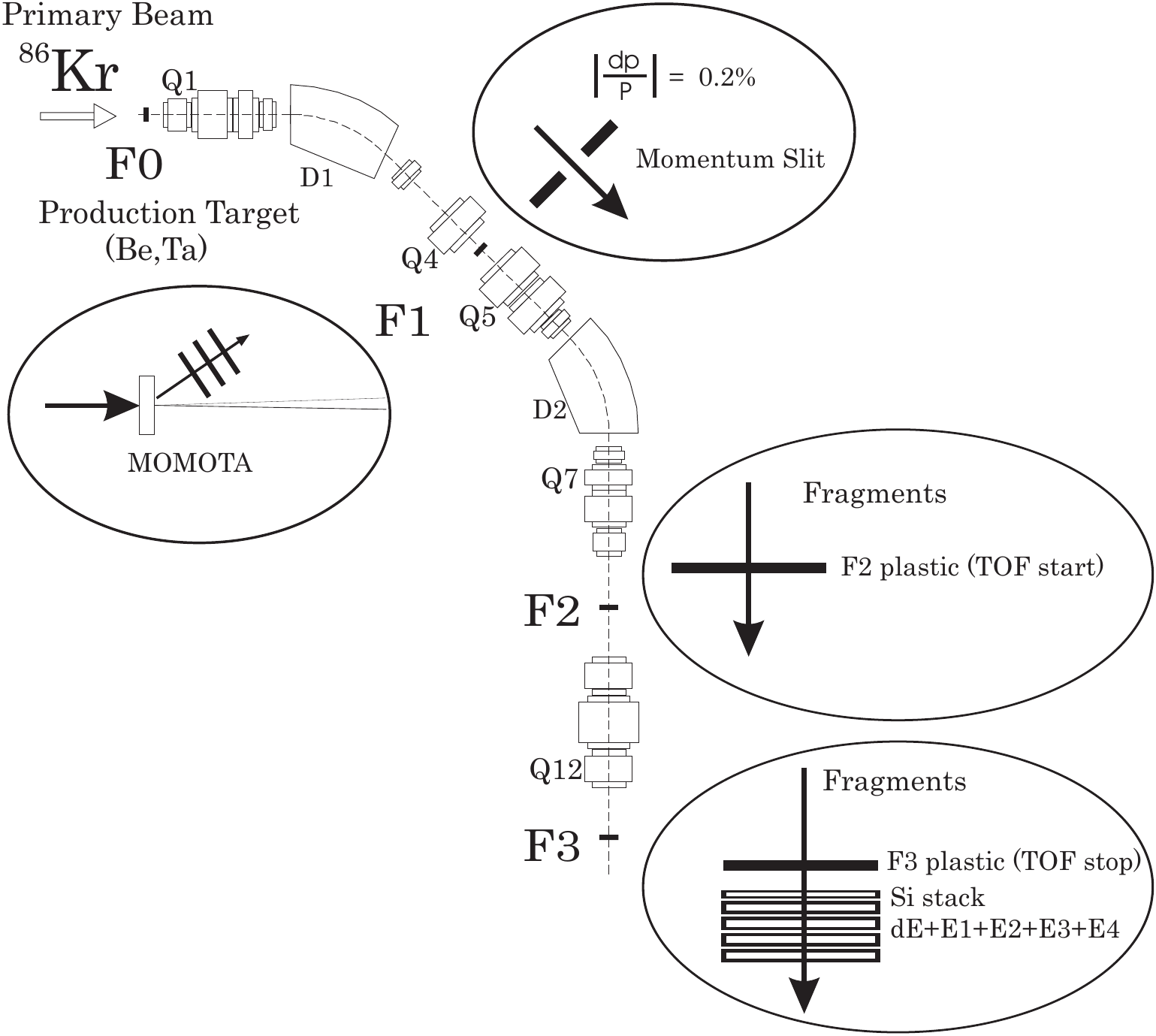}
\caption{RIPS fragment separator consisting of two dipoles (D1 and D2) and twelve quadrupoles (Q1--Q12). The momentum acceptance was determined by the momentum slit placed at F1. The beam intensity monitor (MOMOTA) is shown in the top left oval below the target position. The particle identification setup was located at the F2 and F3 focal planes.}\label{fig2}
\end{figure}

Projectile-like fragments produced in interactions of the primary beam with the target nuclei were collected and identified using the RIPS separator \cite{kub92} located in experimental areas D and E6 as shown in Fig. \ref{fig1}. The schematic layout of RIPS is shown in Fig. \ref{fig2}. The RIPS fragment separator consists of two 45$^\circ$ dipole magnets (D1, D2), and twelve quadrupoles (Q1--Q12). The first section gives a dispersive focus at the F1 focal plane allowing measurement of the magnetic rigidity of the particles. The second stage compensates the dispersion of the first section and gives a double achromatic focus at the F2 focal plane. The quadrupole triplet of the last section produces the third focus at the F3 focal plane, where the main part of the particle identification setup was installed.

All measurements were performed using the RIPS fragment separator in a narrow momentum acceptance mode. The momentum opening, d$p/p$, was limited to 0.2\% using a slit in the dispersive image of the separator, F1 (see the top right oval in Fig. \ref{fig2}). In this configuration, the measured particles have trajectories close to the axis of the fragment separator simplifying the transmission calculations. Furthermore, a narrow momentum acceptance allows measuring the fragment cross sections in the magnetic rigidity between primary beam charge states. The disadvantage is that in order to measure the momentum distributions over a wide range of fragmentation products, we had to take measurements at many different magnetic settings. For reactions with the \nuc{9}{Be} target we covered 1.79--2.93 Tm in 45 steps and for \nuc{181}{Ta} target we scanned the region of 1.79--2.35 Tm in 29 settings. To avoid excessive dead-time in the data acquisition the primary beam intensity was optimized at each magnetic rigidity such that the counting rate of the first silicon PIN detector was approximately 900--1000 counts per second.

\begin{figure}
\includegraphics[width=0.5\textwidth]{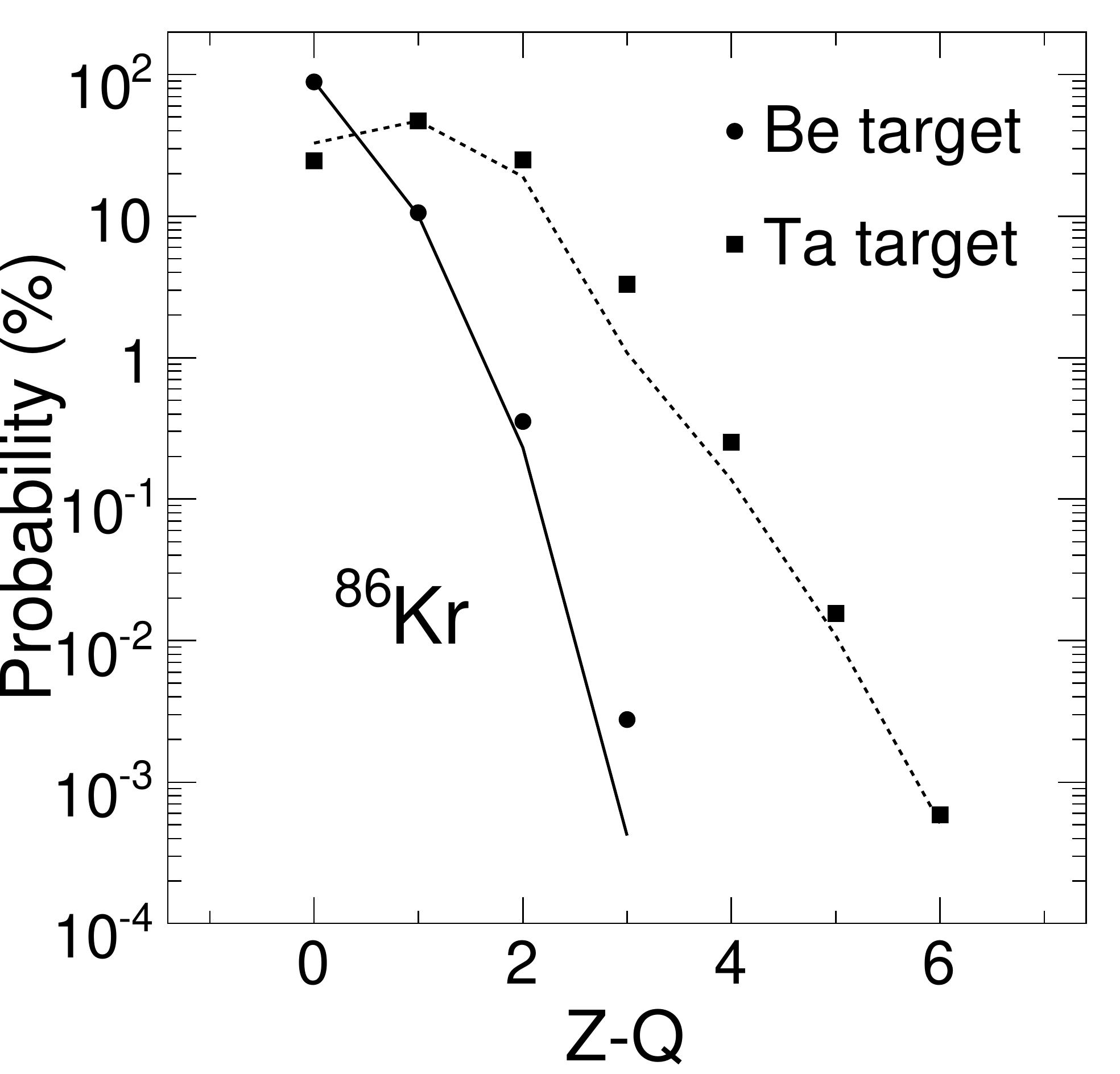}
\caption{Primary beam charge state distributions for \nuc{86}{Kr}+\nuc{9}{Be} (closed circles) and \nuc{86}{Kr}+\nuc{181}{Ta} (closed squares) plotted as a function of number of unstripped electrons, $Z-Q$. Solid and dashed curves show calculation by GLOBAL code \cite{sch98} as implemented in LISE++ \cite{baz02} for \nuc{9}{Be} and \nuc{181}{Ta} targets, respectively.}\label{fig3}
\end{figure}

Fragments with mass number, $A$, proton number, $Z$, and charge state, $Q$, measured in our study ($25\leq Z\leq 36$) were not fully stripped of electrons. However, only the charge state distributions of the \nuc{86}{Kr} primary beam were measured. The measurement was done at the F1 dispersion plane where different charge states of one ion traveling at the same velocity are spatially separated \cite{moc06}. The measured primary charge state probability distributions for \nuc{9}{Be} (filled circles) and \nuc{181}{Ta} (filled squares) targets are plotted in Fig. \ref{fig3} as a function of the number of unstripped electrons, $Z-Q$. Predictions from the charge state distribution code GLOBAL \cite{sch98}, as implemented in LISE++ \cite{baz02}, are shown as solid and dotted lines for \nuc{9}{Be} and \nuc{181}{Ta} targets. The predictions tend to decrease more steeply for the \nuc{86}{Kr}+\nuc{9}{Be} reactions. The overall prediction is quite good considering the fact that the GLOBAL code was developed for heavier projectiles ($Z>53$) at higher energies ($E>100$ \mevu) \cite{sch98}. The measured charge state distribution of the \nuc{86}{Kr} primary beam showed that almost 10\% of the intensity is in the \nucc{86}{Kr}{35+} charge state after passing through the \nuc{9}{Be} target (Fig. \ref{fig3}). The fraction is much larger in the case of \nuc{181}{Ta} target because the charge state distribution is broader.

To properly identify all fragments and their charge states in our analysis, the general $B\rho$-$ToF$-$\Delta E$-$TKE$ \cite{baz90} particle identification technique was used on an event-by-event basis. The magnetic rigidity, $B\rho$, was given by the magnetic setting of the RIPS fragment separator. The time of flight, $ToF$, was measured between F2 and F3 plastic scintillators (see Fig. \ref{fig2}) separated by a flight path of 6 m. The energy loss, $\Delta E$, was measured with a 350 $\mu$m-thick silicon PIN detector. The total kinetic energy, $TKE$, was reconstructed by measuring the energy deposited by the particles in a stack of 5 silicon PIN detectors (labeled $\Delta E$, $E1$, $E2$, $E3$, $E4$ in Fig. \ref{fig2}). 

\begin{figure}
\includegraphics[width=0.49\textwidth]{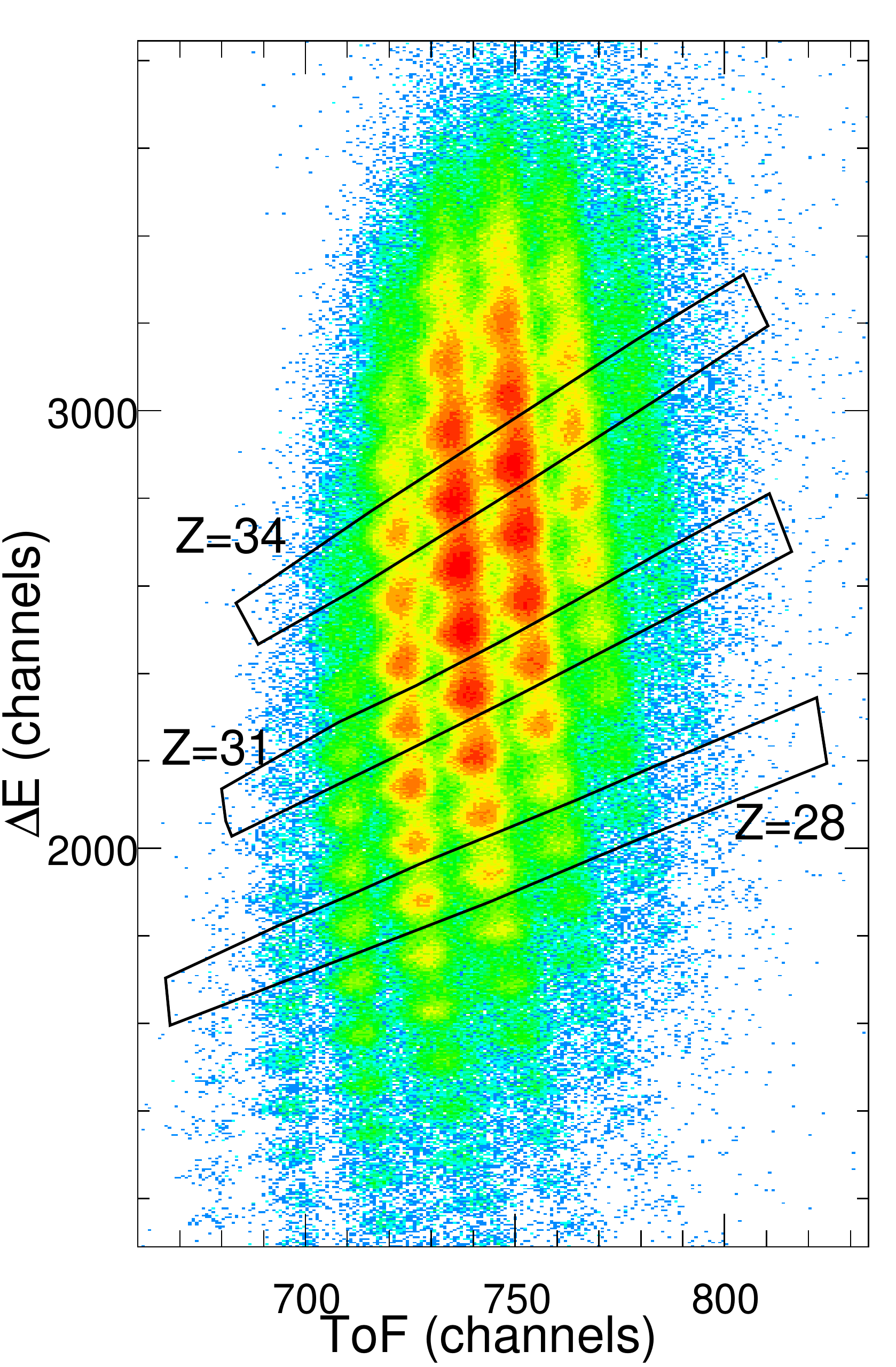}
\includegraphics[width=0.49\textwidth]{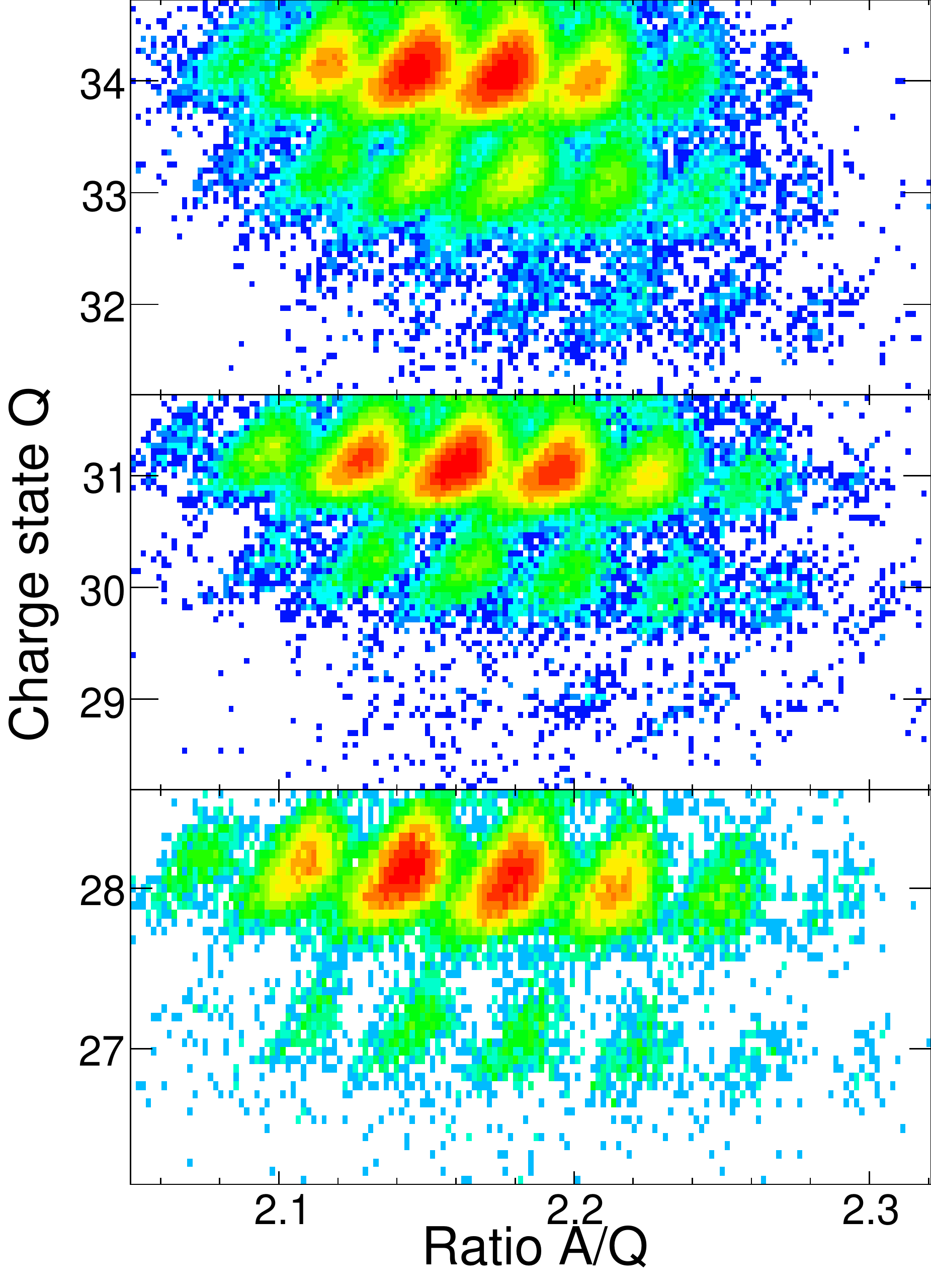}
\caption{(Color online) Particle identification spectrum for the \nuc{86}{Kr}+\nuc{9}{Be} reaction measured at a 2.07 Tm magnetic rigidity setting. Left panel shows the PID with three gates around elements with $Z=28$, 31, 34. Right panel shows the corresponding projections to charge state, $Q$, versus $A/Q$ ratio plane of events within  from bottom to top, respectively.}\label{fig4}
\end{figure}

A typical raw experimental particle identification (PID) plot, $\Delta E$ versus $ToF$, is shown in the left panel of Fig. \ref{fig4}. The identification of individual groups of events was done by recognizing typical features of the PID spectrum and locating a hole corresponding to the particle-unbound \nuc{8}{Be} nucleus \cite{moc06a}. The spectrum for the \nuc{86}{Kr}+\nuc{9}{Be} reaction at $B\rho=2.07$ Tm is shown in Fig. \ref{fig4} and shows 3 gates around elements with $Z=28$, 31, and 34. The right panels display projections of events from these gates to charge state, $Q$, versus ratio $A/Q$ plane. The fully stripped ($Z-Q=0$) and hydrogen-like ($Z-Q=1$) charge states for all 3 selected elements are very well separated. Similar projections were constructed for fragments with $25\leq Z\leq 36$ at all magnetic rigidity settings in our analysis. 

Each experimental run took data for one $B\rho$ setting of the RIPS fragment separator. The number of events, $N(A, Z, Q)$, for a fragment with mass number, $A$, proton number, $Z$, and charge state, $Q$, were extracted from the calibrated PID spectra similar to the one in Fig. \ref{fig4}. The differential cross sections, $\mathrm{d}\sigma/\mathrm{d}p$, were calculated taking into account the number of beam particles, $N_B$, number of target nuclei per square centimeter, $N_T$, live-time ratio, $\tau_{LIVE}$, and the transmission efficiency through the RIPS fragment separator, $\varepsilon$,
\begin{equation}\label{eq1}
\frac{\mathrm{d}\sigma}{\mathrm{d}p}(A,Z,Q)=\frac{N(A,Z,Q)}{N_T N_B \Delta p \tau_{LIVE}}\frac{1}{\varepsilon},
\end{equation}
where $\Delta p$ denotes the momentum opening.

\begin{figure}
\includegraphics[width=0.5\textwidth]{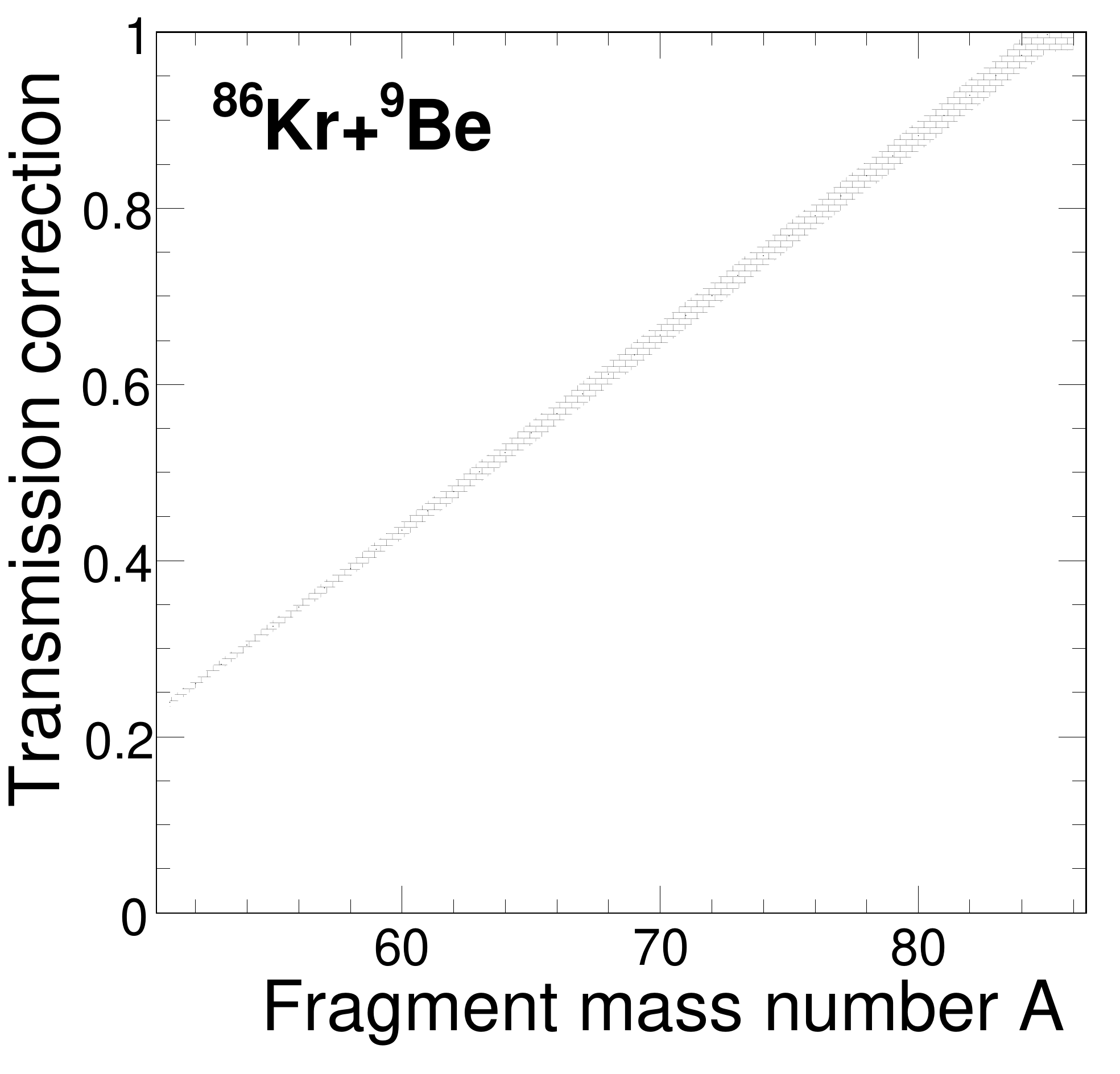}
\caption{Dependence of the transmission correction factor, $\varepsilon$, on fragment mass number, A, for the \nuc{86}{Kr}+\nuc{9}{Be} reactions.}\label{fig5}
\end{figure}

The transmission efficiency correction, $\varepsilon$, is assumed to be factorized into two independent components: momentum corrections and angular corrections. Momentum corrections take into account the loss of fragments caused by the momentum slit at the F1 focal plane. This effect is independent of fragment species and the $B\rho$ setting. A correction value of $98\pm 2$\% was obtained from simulations using a universal Monte Carlo ion optics code MOCADI \cite{iwa97}. The angular corrections account for a finite angular acceptance of the RIPS fragment separator in the perpendicular (transverse) plane with respect to the beam direction. Since the current experiment does not measure the momentum in the transverse direction, we modeled the width of the momentum distribution of a fragment with a mass number, $A$, by a Gaussian distribution with variance, $\sigma_{\bot}$, prescribed in ref. \cite{bib79}:
\begin{equation}\label{eq2}
\sigma^{2}_{\bot}=\sigma^2_0\frac{A(A_P-A)}{A_P-1}+\sigma^2_D\frac{A(A-1)}{A_P(A_P-1)},
\end{equation}
where $A_P$ is the mass number of the projectile and $\sigma_D$ is the orbital dispersion. The first term in Eq. (\ref{eq2}) comes from the Goldhaber model \cite{gold74}, which describes the width of longitudinal momentum distribution of fragments produced at high projectile energies. The value of $\sigma_0$ was determined by fitting the experimental longitudinal distributions. Values of $147\pm 5$ and $153\pm 5$ \mevc\, were obtained for reactions with \nuc{9}{Be} and \nuc{181}{Ta} targets, respectively. The second term in Eq. (\ref{eq2}) takes into account the deflection of the projectile by the target nucleus \cite{day86} and is significant only for fragments with masses close to the projectile and at low and intermediate beam energies. We estimated the $\sigma_D$ parameter to be $225\pm 25$ \mevc\, for both investigated reactions, based on the \nuc{16}{O} fragmentation data measured at 90 \mevu\, \cite{bib79}. Portions of the Gaussian angular distributions transmitted through the RIPS fragment separator define the angular transmission and were calculated using LISE++ \cite{baz02} and verified with MOCADI simulations \cite{moc06}. The transmission correction, $\varepsilon$, consisting of the product of the angular and momentum corrections is plotted in Fig. \ref{fig5} for the \nuc{86}{Kr}+\nuc{9}{Be} reaction. The final transmission correction, $\varepsilon$, varies from 0.98 for fragments close to the projectile to approximately 0.25 for the lightest fragments in our analysis ($A\approx 50$). The transmission correction for the \nuc{86}{Kr}+\nuc{181}{Ta} reaction is very similar to the one shown in Fig. \ref{fig5}. 

In our fragmentation measurements the beam intensity varied between $10^6$ and $10^{11}$ pps. The beam intensity was monitored by a telescope located at approximately 60$^{\circ}$ with respect to the beam direction and approximately 25 cm from the target. The top left oval in Fig. \ref{fig2} shows a schematic drawing of the monitor (MOMOTA) at the target position. The monitor consists of three plastic scintillators and detects the light particles produced in nuclear reactions in the production target. Only triple coincidence rates were considered as valid signals. Since the production of light particles depends on the reaction of beam and target nuclei, the monitor rates must be calibrated to the beam intensity for each reaction system studied. Unfortunately, we could not use the Faraday Cup (FC) to calibrate the beam intensity.  The FC was located approximately 5 cm downstream from the target position and the monitor reading was affected by the particles scattered off the FC during the primary beam intensity calibration. To obtain an absolute calibration of the monitor, direct rates of \nucc{86}{Kr}{33+} and \nucc{86}{Kr}{31+} particles for the \nuc{9}{Be} and \nuc{181}{Ta} targets, respectively, were measured at the F2 focal plane using the plastic scintillator. The statistical uncertainties of these measurements were less than 5\%. From Fig. \ref{fig3}, probabilities of \nucc{86}{Kr}{33+} and \nucc{86}{Kr}{31+} charge states are found to be 0.0028\% and 0.016\%, respectively. This allowed us to calculate the primary beam intensity for these two measurements, thus establishing absolute beam intensity calibration points for the \nuc{9}{Be} and \nuc{181}{Ta} targets. The linearity (better than 1\%) in the beam intensity range used in our experiments for the monitor telescope was confirmed by measuring the fragment flux with different F1 slit openings. 

\section{Momentum distributions}
\begin{figure}
\includegraphics[width=0.5\textwidth]{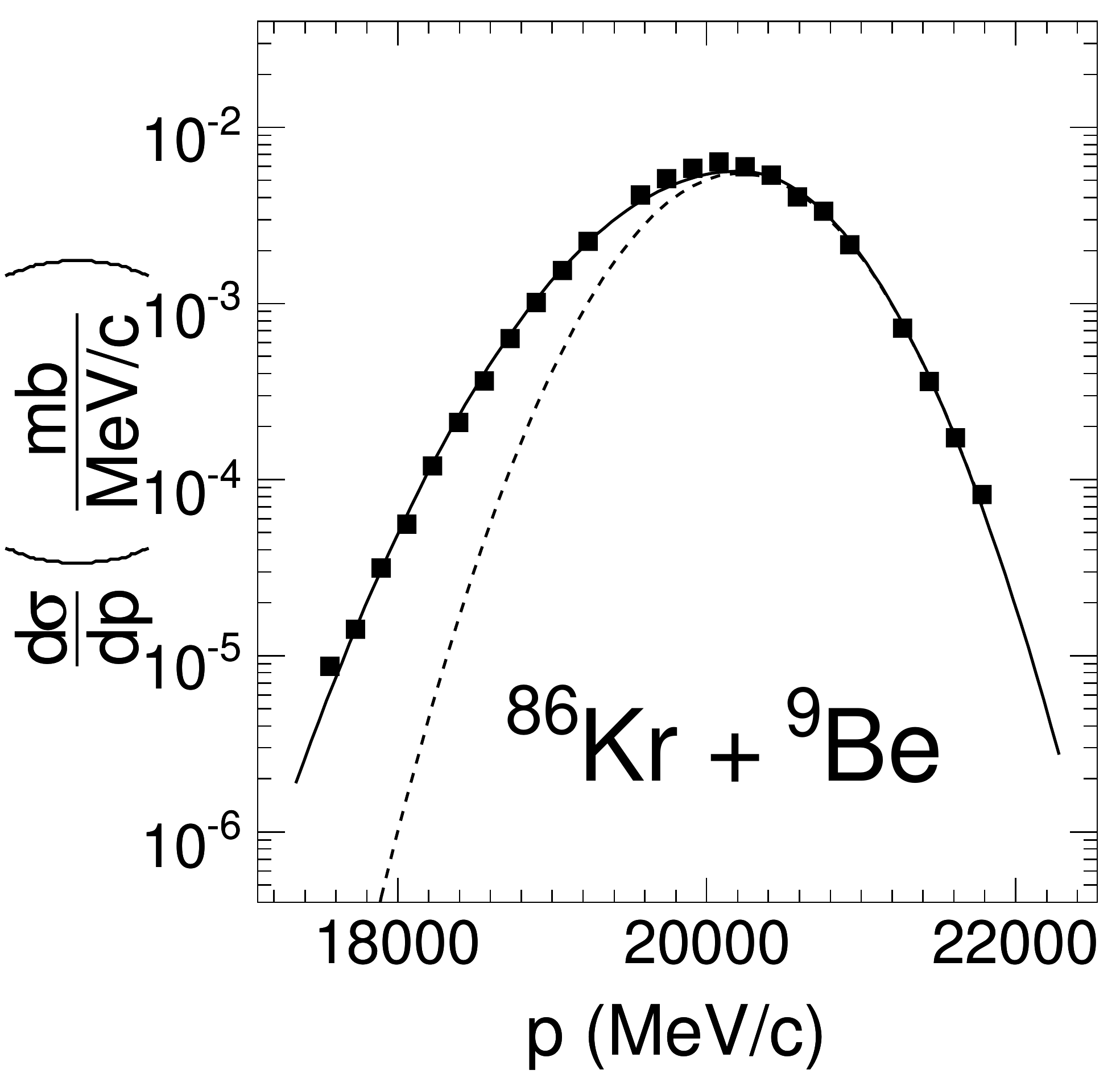}
\caption{Momentum distributions for \nucc{64}{Zn}{30+} produced in fragmentation of \nuc{86}{Kr} on the \nuc{9}{Be} target. The solid curve represents a fit with Eq. (\ref{eq4}) and the dotted curve is a Gaussian fit, to the right side of the momentum distribution, to show the asymmetry of the experimental distribution.}\label{fig6}
\end{figure}

The fragment momentum distributions were obtained by plotting individual differential cross sections as a function of measured momentum (calculated from the magnetic rigidity, $B\rho$) for all fragments and their charge states. 
The momentum distributions obtained from projectile fragmentation at intermediate energy are asymmetric \cite{baz90,moc06}. Fig. \ref{fig6} displays a typical momentum distribution in our analysis for \nucc{64}{Zn}{30+}. The dashed curve represents a fit with a single Gaussian function. As the distributions have low momentum tails, we fit the data with the following function \cite{not07,moc06}:
\begin{equation}\label{eq4}
\frac{d\sigma}{dp}= \left\{
\begin{array}{ll}
 S\cdot \exp\left(-(p-p_0)^2/(2\sigma_L^2)\right) & \textrm{for $p\leq p_0$},\\
 S\cdot \exp\left(-(p-p_0)^2/(2\sigma_R^2)\right) & \textrm{for $p>p_0$},\\
\end{array}
\right.
\end{equation}
where $S$ is the normalization factor, $p_0$, is the peak position of the distribution, and $\sigma_L$ and $\sigma_R$ are widths of ``left'' and ``right'' halves of two Gaussian distributions used to fit the momentum distributions. The solid curves in Fig. \ref{fig6} are the best fits obtained by minimization of $\chi^2$ using Eq. (\ref{eq4}). For most fragments we observe very good agreement between the data and the fit over three orders of magnitude.

\section{Cross-section measurements}
The cross section of a fragment in a given charge state was determined by integrating the area of its momentum distribution. For fragments with well-measured momentum distributions, such as the one shown in Fig. \ref{fig6}, the cross sections were extracted from fitting the momentum distributions using Eq. (\ref{eq4}). However, approximately 40\% of the measured fragments had incomplete momentum distributions that may consist of only a few points near the top of the peak.  For these fragments, we used the systematics of $p_0$, $\sigma_L$, and $\sigma_R$ obtained from fragments with complete momentum distributions to calculate the cross sections with function in Eq. (\ref{eq4}).

At 64 \mevu, the fragment yield is distributed over different charge states.  The total fragmentation cross sections are obtained by summing these contributions. For the \nuc{86}{Kr}+\nuc{9}{Be} reaction system we analyzed fully stripped fragments with $Z-Q=0$ charge states and corrected the final fragment cross sections using charge state distributions predicted by GLOBAL. The calculated corrections vary between 1--9\% for $25\leq Z\leq 36$ isotopes. For the \nuc{86}{Kr}+\nuc{181}{Ta} reaction we sum the cross sections of the 3 most abundant charge states ($Z-Q=0$, 1, 2) to harvest most of the cross section. Corrections for fragment cross sections using GLOBAL vary between 0.1--3\% for $25\leq Z\leq 36$ isotopes.

\begin{figure}
\includegraphics[width=0.5\textwidth]{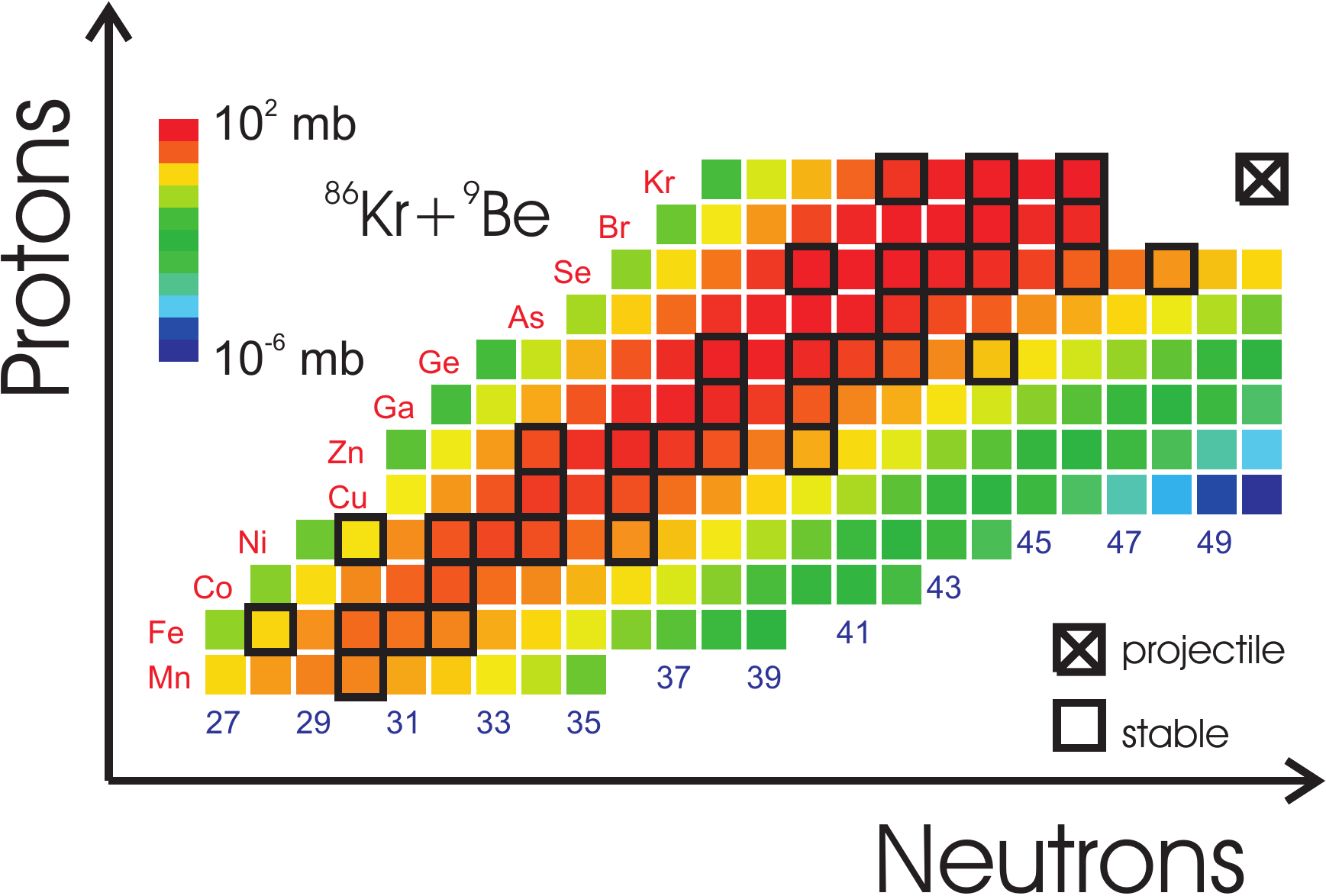}
\caption{(Color online) Measured cross sections for 180 fragments produced in the \nuc{86}{Kr}+\nuc{9}{Be} reactions.}\label{fig7}
\end{figure}

For fragments with complete momentum distributions, uncertainties in the fragmentation cross sections of 7--12\%, were calculated based on the statistical uncertainty, the beam intensity calibration (5\%), the errors from the fitting procedure and the transmission uncertainty (2--8\%). For fragments measured with incomplete momentum distributions, additional systematic errors stemming from the extrapolation of the parameters of $p_0$, $\sigma_L$, and $\sigma_R$ were included.
An overall view of the fragment cross sections for the \nuc{86}{Kr}+\nuc{9}{Be} reaction system in the style of the nuclear chart, is shown in Fig. \ref{fig7}. The range of the measured cross sections spans over 9 orders of magnitude, from $15\pm 7$ pb (\nuc{79}{Cu}) to $38\pm 4$ mb (\nuc{82}{Kr}).

\section{Cross-section results}
\begin{figure}
\includegraphics[width=0.9\textwidth]{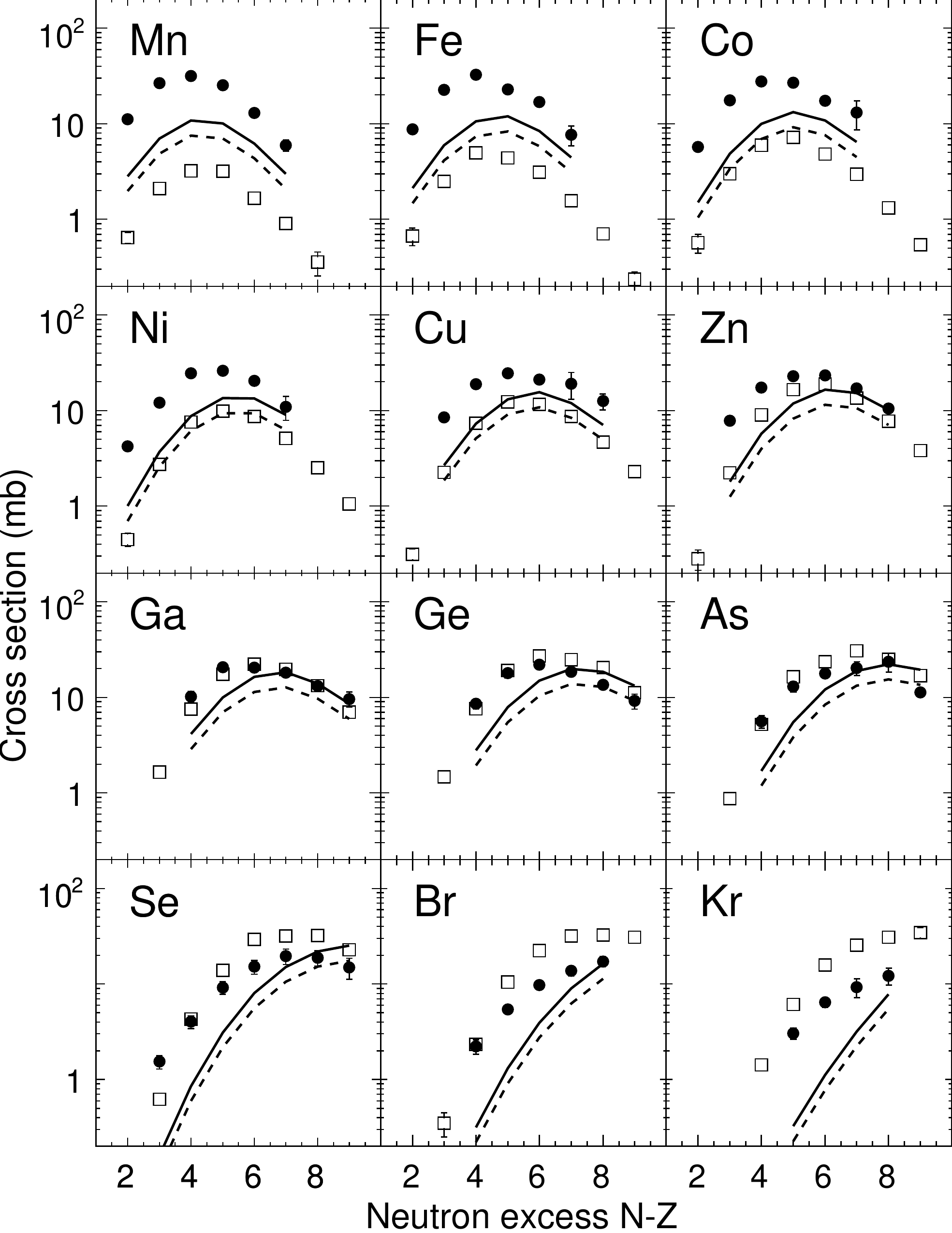}
\caption{Measured cross sections presented as isotope distributions for $25\leq Z\leq 36$ elements detected in the \nuc{86}{Kr}+\nuc{181}{Ta} reactions (filled circles) and in the \nuc{86}{Kr}+\nuc{9}{Be} reactions (open squares) at 64 \mevu. EPAX calculations are shown as dashed (\nuc{86}{Kr}+\nuc{9}{Be}) and solid (\nuc{86}{Kr}+\nuc{181}{Ta}) curves.}\label{fig8}
\end{figure}

\begin{figure}
\includegraphics[width=0.9\textwidth]{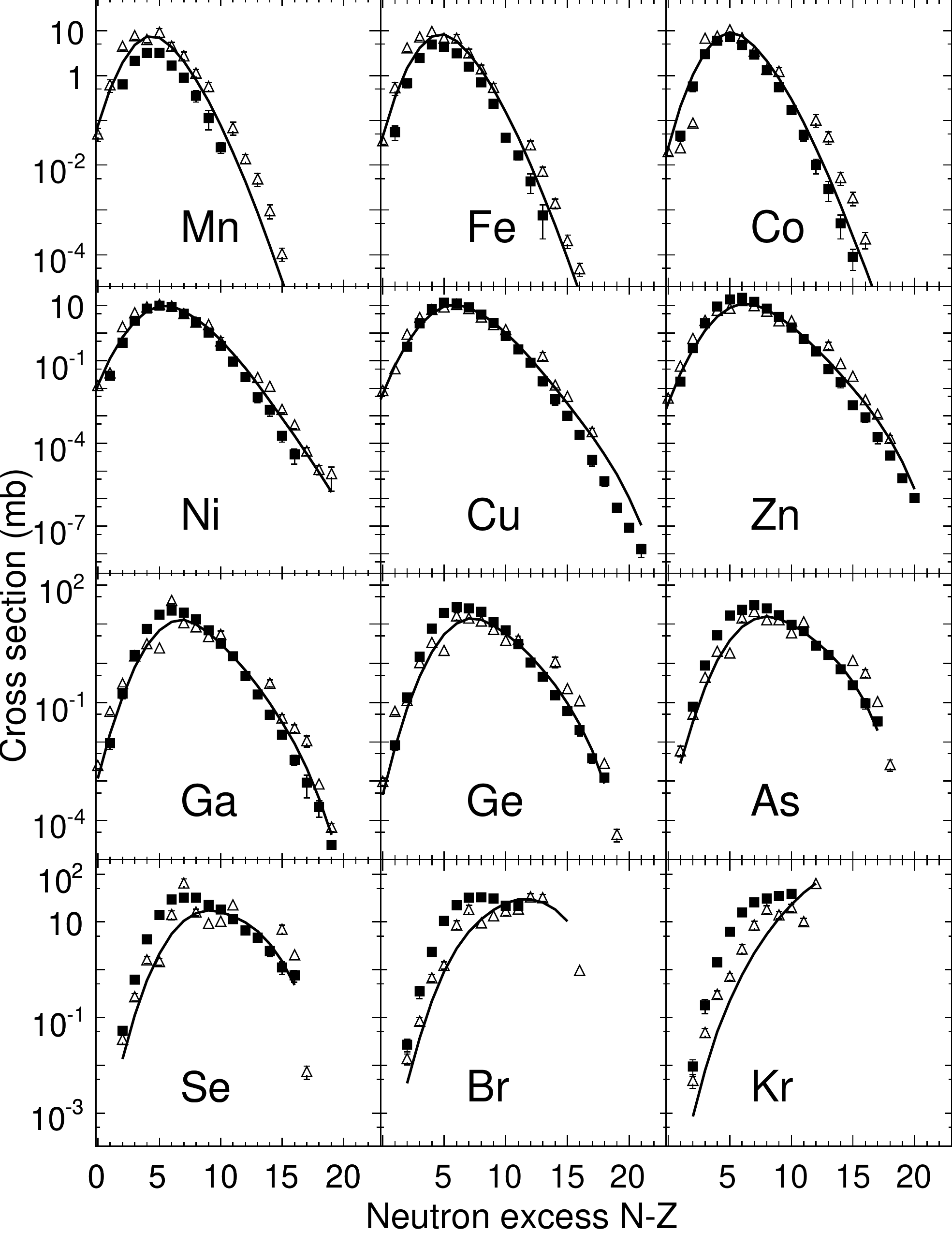}
\caption{Measured cross sections presented as isotope distributions for $25\leq Z\leq 36$ elements detected in the \nuc{86}{Kr}+\nuc{9}{Be} reactions at 64 \mevu. Experimental fragmentation data are shown as filled squares. EPAX predictions are shown as solid curves. For comparison, open triangles show the published data of \nuc{86}{Kr}+\nuc{9}{Be} at 500 \mevu\, \cite{web94}.}\label{fig9}
\end{figure}

Fig. \ref{fig8} shows the cross sections for fragments extracted from the \nuc{86}{Kr}+\nuc{181}{Ta} analysis as closed circles. Each panel represents isotope cross-section data for one element ($25\leq Z\leq 36$), plotted as a function of neutron excess, $N-Z$, of each isotope. For the \nuc{86}{Kr}+\nuc{181}{Ta} reaction system, interference from the many charge states of the beam limits the span of measured fragments for each element. Our requirement, that the three most abundant charge states should have quantifiable counts above background in the analysis further reduced the number of data points to 70 isotopes for the \nuc{86}{Kr}+\nuc{181}{Ta} system. In contrast, cross sections for 180 isotopes were obtained for the \nuc{86}{Kr}+\nuc{9}{Be} system as shown in Fig. \ref{fig9}.

\begin{figure}
\includegraphics[width=0.5\textwidth]{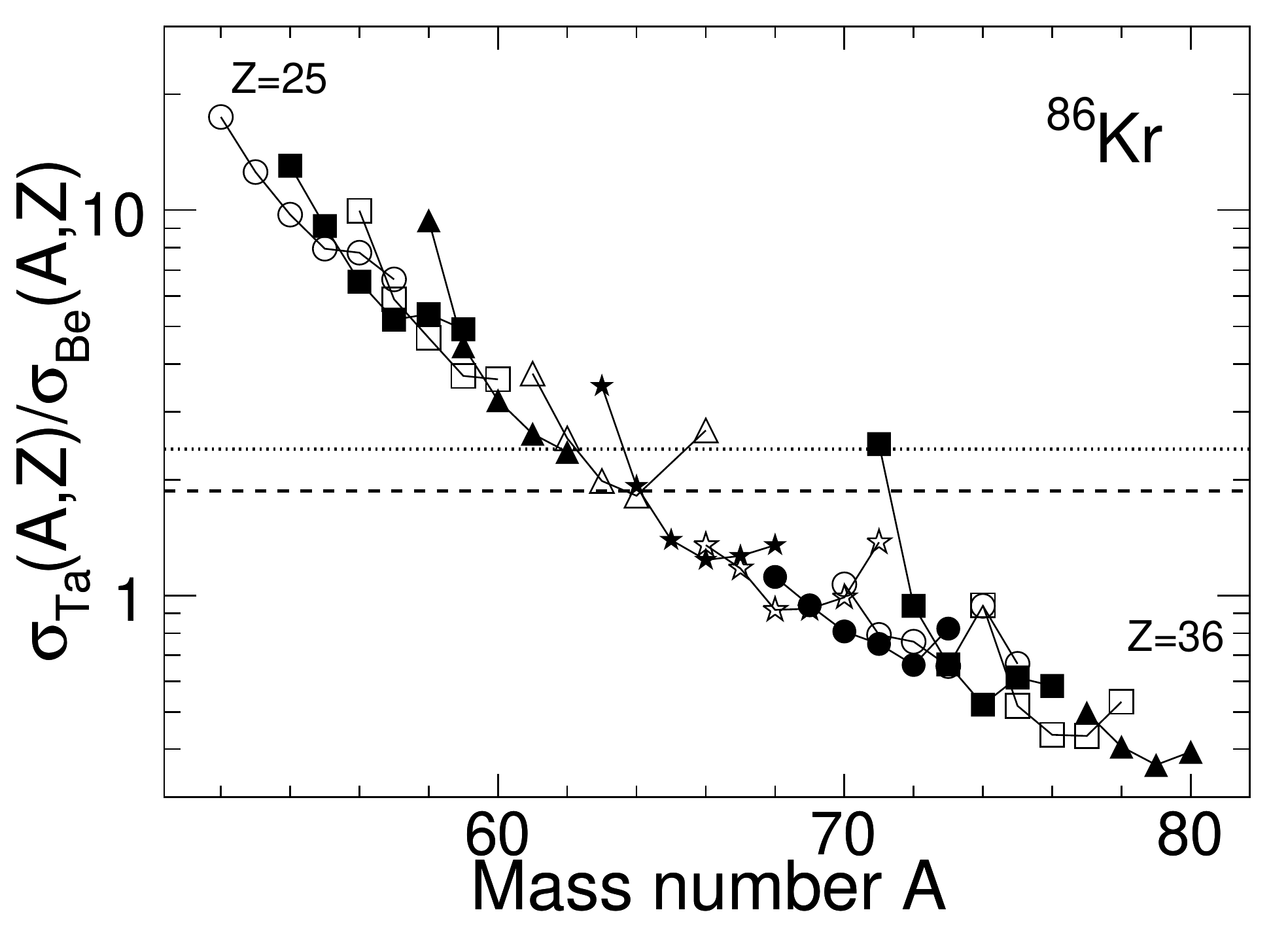}
\caption{Ratios of the fragmentation cross sections on Ta and Be targets, $\sigma_{\mathrm{Ta}}(A,Z)/ \sigma_{\mathrm{Be}}(A,Z)$, for fragments with $25\leq Z\leq 36$ for the \nuc{86}{Kr} beam. Only ratios with relative errors smaller than 25\% are shown.  Open and solid symbols represent odd and even elements starting with $Z=25$. The horizontal dashed and dotted lines indicate the ratio calculated by the EPAX formula and Eq. (\ref{eq5}), respectively.}\label{fig10}
\end{figure}

For comparison, fragment cross sections for the \nuc{86}{Kr}+\nuc{9}{Be} reactions are plotted as open squares in Fig. \ref{fig8}. More light fragments are produced in the projectile fragmentation of the \nuc{86}{Kr} nuclei with \nuc{181}{Ta} than \nuc{9}{Be} targets. This increase is seen clearly in Fig. \ref{fig10} where the ratios of isotope yields from the two different targets, $\sigma_{\mathrm{Ta}}(A,Z)/\sigma_{\mathrm{Be}}(A,Z)$, are plotted as a function of fragment mass number, $A$, and $\sigma_{\mathrm{Ta}}(A,Z)$ and  $\sigma_{\mathrm{Be}}(A,Z)$ denote cross sections of an isotope $(A,Z)$ measured with \nuc{181}{Ta} and \nuc{9}{Be} targets, respectively.  For clarity of the presentation, only the target isotope ratios with relative errors smaller than 25\% are shown. Elements with odd and even $Z$’s are represented by open and closed symbols, respectively, with the open circles starting at $A\approx 52$ representing the Mn isotopes and the solid triangles near $A\approx 80$ denoting the Kr isotopes.  Within an element (data points with same symbol), there seems to be an increase in the fragment cross sections from reactions with Ta targets for both very neutron-rich and proton-rich isotopes. The trend is not as clear here due to the limited range of isotopes measured in the \nuc{86}{Kr}+\nuc{181}{Ta} reactions. (Similar trends have been observed in the projectile fragmentation of \nuc{40,48}Ca and \nuc{58,64}Ni isotopes \cite{moc06}.)  The experimental target isotope ratios, $\sigma_{\mathrm{Ta}}(A,Z)/\sigma_{\mathrm{Be}}(A,Z)$, exhibit an overall increase with decreasing fragment mass in Fig. \ref{fig10}. For fragments lighter than $A\approx 50$, the enhancement exceeds a factor of 10.  Such dependence is not expected in the limiting fragmentation model. In the geometrical limit the cross sections are proportional to the sum of nuclear radii squared \cite{kox84}, so the target isotope ratios are given by:
\begin{equation}\label{eq5}
\frac{\sigma_{\mathrm{Ta}}(A,Z)}{\sigma_{\mathrm{Be}}(A,Z)}=\frac{\left( A_{\mathrm{Kr}}^{1/3}+A_{\mathrm{Ta}}^{1/3}\right)^2}{\left( A_{\mathrm{Kr}}^{1/3}+A_{\mathrm{Be}}^{1/3}\right)^2}=2.4,
\end{equation}
where $A_{\mathrm{Kr}}=86$, $A_{\mathrm{Ta}}=181$, and $A_{\mathrm{Be}}=9$. This limit is shown as a dotted line in the figure.  In the EPAX formula the fragmentation cross section is proportional to the sum of nuclear radii, which stems from the assumption that fragmentation is dominated by peripheral events: 
\begin{equation}\label{eq6}
\frac{\sigma_{\mathrm{Ta}}(A,Z)}{\sigma_{\mathrm{Be}}(A,Z)}=\frac{\left( A_{\mathrm{Kr}}^{1/3}+A_{\mathrm{Ta}}^{1/3}-2.38\right)}{\left( A_{\mathrm{Kr}}^{1/3}+A_{\mathrm{Be}}^{1/3}-2.38\right)}=1.9.
\end{equation}
This EPAX limit is shown as a dashed line in the figure. The cross-section enhancement trends suggest that light, rare isotopes may be produced more abundantly using a heavy target such as \nuc{181}{Ta}.  However, one must keep in mind the large difference in atomic mass of the two target materials (approximately a factor of 20). To compensate for the low atomic density in Ta or similar targets, thick foils must be used, and effects such as the broad charge state distribution for heavy targets, the energy loss, and angular straggling must be taken into account. However, if the rising trend of the target isotope ratios for the \nuc{86}{Kr} primary beam continues for light isotopes, heavy targets such as Ta may be a better choice for the production of light neutron-rich and proton-rich isotopes close to the drip lines \cite{sak99}.

For both investigated systems, we also observed differences between the EPAX calculated and observed maxima of the isotopic distribution for elements close to the projectile (Ge--Kr). A similar systematic discrepancy between the intermediate energy fragmentation data and EPAX parameterization has been reported before \cite{sum03, moc06}. The Fermi spheres of the target and projectile nuclei have larger overlap at intermediate energies than at relativistic energies. There may be increasing contributions to the prefragments with charge numbers greater than that of the projectile from the transfer-type reactions. Subsequent decay of these primary fragments feeds the less neutron-rich isotopes close to the projectile.

\begin{figure}
\includegraphics[width=\textwidth]{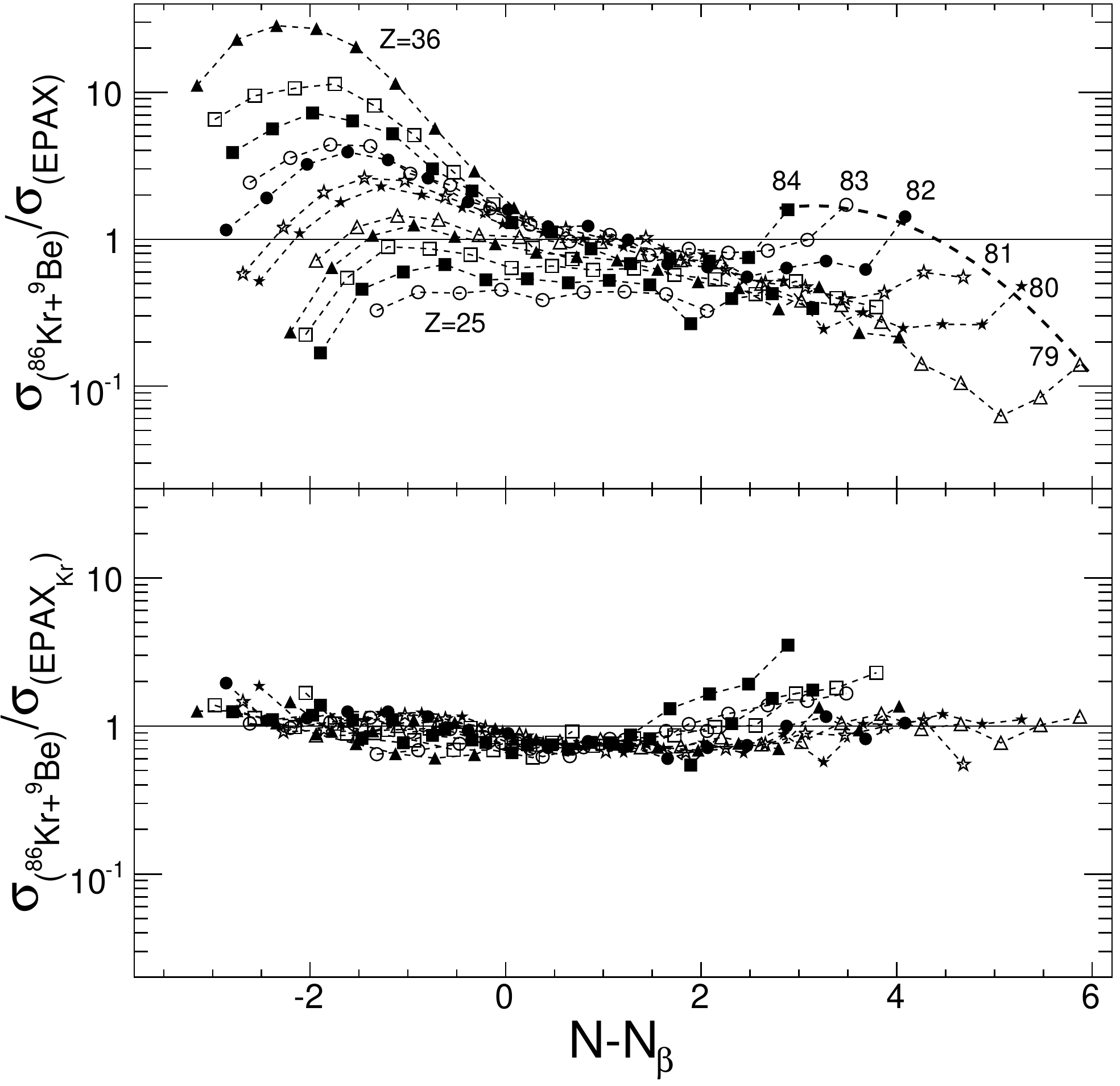}
\caption{Ratio of the experimental cross sections and predicted cross sections from EPAX (top panel) and our modified $\mathrm{EPAX_{Kr}}$ formula (bottom panel). For clarity, isotopes from each element are joined by the dashed lines. Open and solid symbols represent odd and even elements from $Z=25$ to $Z=36$. The bold dashed curve joining the $N=50$ proton-removed isotopes (\nuc{84}{Se}, \nuc{83}{As}, \nuc{82}{Ge}, \nuc{81}{Ga}, \nuc{80}{Zn}, and \nuc{79}{Cu} are labeled with mass number) is obtained from a fit. The curve allows extrapolation of the production estimates of very neutron-rich nucleus such as \nuc{78}{Ni}.}\label{fig11}
\end{figure}

The parameters used in EPAX were obtained by fitting several data sets, including the fragmentation data of \nuc{86}{Kr}+\nuc{9}{Be} at 500 \mevu\, \cite{web94}. For comparison, the latter set of data was plotted as open triangles in Fig. \ref{fig9}, and our data are plotted as closed squares.  There are considerable scatters in the Weber \textit{et al.} data (especially for Ga to Se elements). The cross sections at the peak of the isotopic distributions for Co to Zn elements agree rather well. However, the 500-\mevu\, isotope distributions are wider. These may account for the larger widths from the calculated isotope distributions by EPAX. 
It has been known that EPAX over-predicts the production of very neutron-rich fragments \cite{moc06a, not07}. The top panel of Fig. \ref{fig11} shows the ratio of the measured cross sections divided by the EPAX predictions as a function of the neutron number from the neutron stability line, $N_{\beta}$. For convenience, we adopt the same stability line for a chain of isobars, $A$, as used in EPAX \cite{sum00}:
\begin{equation}\label{eq7}
N_{\beta}=A-\frac{A}{1.98-0.0155 A^{2/3}}.
\end{equation}
Other choices of the stability line lead to the same conclusions. The same convention of the symbols used in Fig. \ref{fig10} is adopted here with the open circles (top left corner in Fig. \ref{fig10}) denoting $Z=25$ isotopes and closed triangles (lower right corner in Fig. \ref{fig10}) denote $Z=36$ isotopes. EPAX predicts isotopes near the stability line to better than a factor of 2. However, starting around two neutrons beyond the EPAX stability line, over-prediction from EPAX worsens with increasing neutron richness for a fixed element. By extrapolating the proton-removed isotopes ($N=50$) from the \nuc{86}{Kr} projectiles (the right-most points joined by the dashed curve), the over-prediction of the rare neutron-rich nuclei such as \nuc{78}{Ni} could be a factor of 100. 

\begin{figure}
\includegraphics[width=\textwidth]{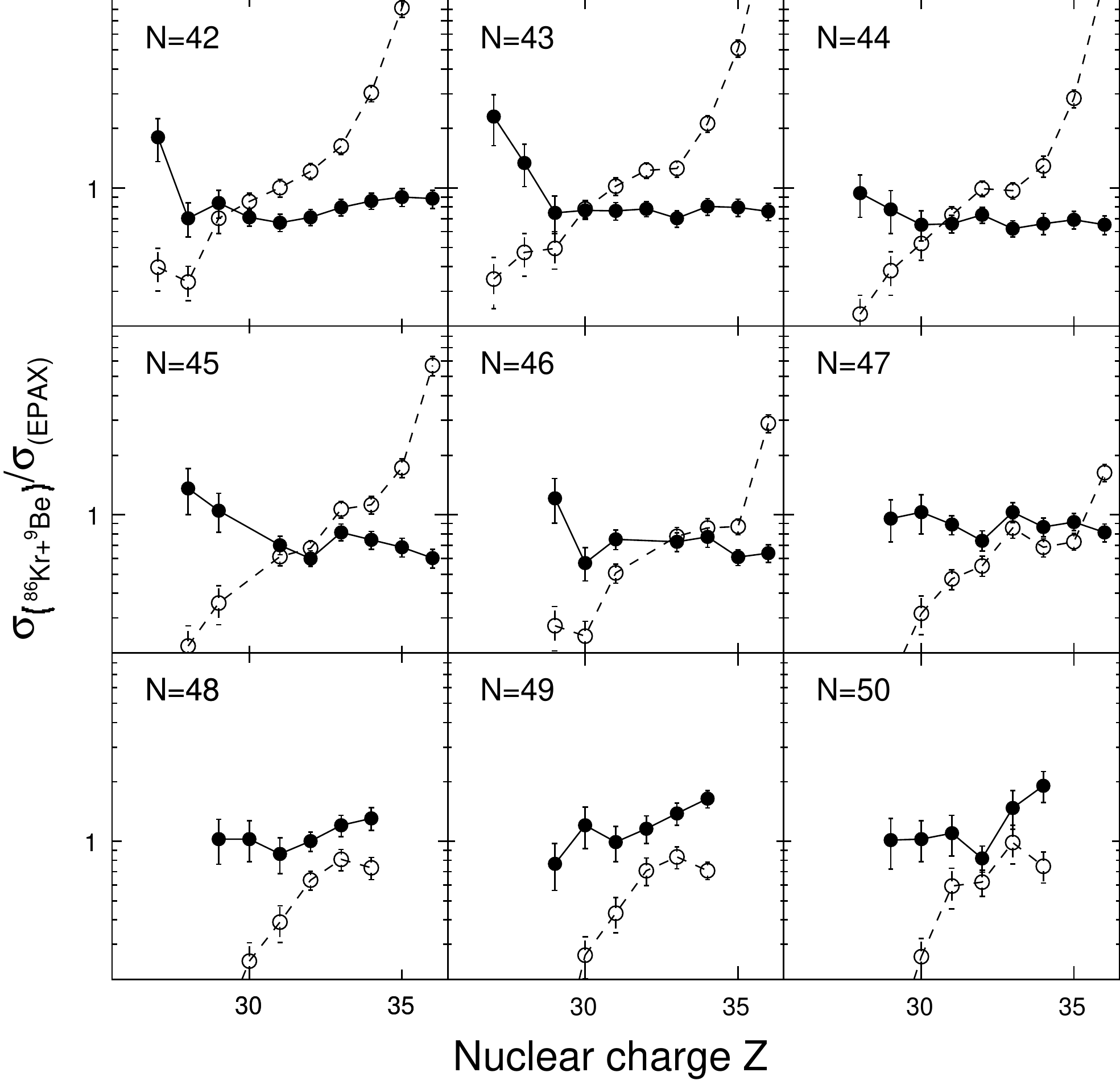}
\caption{Ratios of \nuc{86}{Kr}+\nuc{9}{Be} fragment experimental cross sections to EPAX \cite{sum00} (open symbols) and to our modified $\mathrm{EPAX_{Kr}}$ (solid symbols) predictions plotted as a function of nuclear charge, $Z$, for $42\leq N\leq 50$ isotones.}\label{fig12}
\end{figure}

To examine the behavior of EPAX predictions with respect to neutron-rich nuclei, we plot the ratios of $\sigma_{(\nuc{86}{Kr}+\nuc{9}{Be})}/\sigma_{\mathrm{EPAX}}$ as a function of the atomic number of the fragments for $42\leq N\leq 50$ isotones in Fig. \ref{fig12}. The open circles represent predictions from the standard EPAX calculations. In each panel, the neutron-rich isotopes are those with lowest $Z$. Aside from the pick-up reactions, the most neutron-rich fragments created in the projectile fragmentation reactions of \nuc{86}{Kr} are isotones with $N=50$ (lower right panel). In most cases, the last data point with lowest $Z$ in each isotone chain is only a couple proton numbers away from the most neutron-rich known nuclei. Thus, EPAX predictions on the production of very proton rich and neutron rich isotopes can be off by more than an order of magnitude. Since neutron-rich nuclei are of interest to a variety of problems in astrophysics and nuclear structure the demand for such beams is high. Unfortunately, the inaccuracy in the beam rate estimation using EPAX presents large uncertainties in designing experiments involving these rare isotopes. 

Since the EPAX parameters result from fitting the projectile fragmentation data of \nuc{40}{Ar}, \nuc{48}{Ca}, \nuc{58}{Ni}, \nuc{86}{Kr}, \nuc{129}{Xe}, and \nuc{208}{Pb} with the beam energy above 200-\mevu\, heavy-ion data, better fitting parameters may be obtained if only the present data set is used. The new set of fitting parameters may allow more accurate extrapolation to the yields of very neutron rich nuclei. 

In the original version of EPAX, as briefly described in Appendix A, a total of 24 fitting parameters was obtained. Table \ref{table} lists the parameters used in the original EPAX as well as the modified EPAX parameters used to fit the present data.  (For convenience, we label the EPAX calculations using the new set of parameters $\mathrm{EPAX_{Kr}}$.) The bottom panel of Fig. \ref{fig11} shows the ratio of data over the predictions from $\mathrm{EPAX_{Kr}}$. Compared to the top panel, the overall agreement with the experimental data is much better. This is not surprising considering $\mathrm{EPAX_{Kr}}$ is not a global fit and describes the cross sections for only one reaction. To study how the extrapolations would behave in the neutron-rich region, the new ratios of data over the predictions of $\mathrm{EPAX_{Kr}}$ are plotted as closed points in Fig. \ref{fig12}. Contrary to the ratios using original EPAX parameters, the new ratios are less than a factor of two over a large $Z$ range. However, the behavior of the most neutron rich nuclei ratios do not exhibit a predictable dependence on $Z$. Thus accurate extrapolation to the unmeasured neutron-rich region (the left side of each panel with smaller $Z$ for fixed $N$) cannot be obtained. This could be due to the fact that EPAX is a fitting code that does not include the properties of exotic nuclei such as binding energy or neutron separation energy \cite{fri03} . Better extrapolations will require the use of models that include more physics. However, discussions of such models are beyond the scope of this paper.

\section{Summary}
Fragmentation production cross sections have been measured for \nuc{86}{Kr} primary beam on \nuc{9}{Be} and \nuc{181}{Ta} reaction targets at 64 \mevu. The cross-section ratios obtained with the \nuc{181}{Ta} and \nuc{9}{Be} targets show a fragment mass and charge dependence, contrary to the simple geometrical models. The isotopic distributions of fragments produced in \nuc{86}{Kr}+\nuc{9}{Be} reactions are narrower than those calculated by the EPAX formula resulting in severe cross-section over-predictions for the very neutron-rich isotopes. The availability of comprehensive data, such as those presented here, suggests that it is difficult to extrapolate accurately the cross sections of exotic neutron-rich nuclei with different EPAX fitting parameters \cite{not07, sum06, sun07}. Away from the stability, properties of the exotic nuclei become important, and EPAX does not include basic nuclear property information such as the binding energy.

\begin{acknowledgments}
We would like to thank the operation group of Riken for producing high quality and high intensity \nuc{86}{Kr} beam during our experiment. We thank Dr. K. S\"ummerer for giving us invaluable insights on fitting the EPAX parameters. This work is supported by the National Science Foundation under Grant Nos. PHY-01-10253, PHY-0606007 INT-0218329, and OISE*-0089581.  
\end{acknowledgments}

\appendix*
\section{EPAX parameterization}   
In the EPAX parameterization \cite{sum00} the fragmentation cross section of a fragment with mass, $A$, and nuclear charge, $Z$, created from projectile ($A_p$, $Z_p$) colliding with a target ($A_t$, $Z_t$) is given by:
\begin{equation}\label{eq:epax}
\sigma(A,Z)=Y_An\exp{\left( -R|Z_{prob}-Z|^{U_{\mathrm{n(p)}}}\right)}.
\end{equation}
The first term $Y_A$ describes the sum of the isobaric cross sections with $A$. The second term, $\exp{\left( -R|Z_{prob}-Z|^{U_{\mathrm{n(p)}}}\right)}$, is called the ``charge dispersion,'' and describes the distribution of the elemental cross sections around the maximum value, $Z_{prob}$, for a given mass. The shape of the charge distribution is controlled by the width parameter, $R$, and the exponents, $U_{\mathrm{n}}$ and $U_{\mathrm{p}}$, describe the neutron-rich (n) and proton-rich (p) side, respectively. The neutron-rich fragments are defined with $Z_{prob}-Z>0$ and all others are considered proton-rich. The factor $n=\sqrt{R/\pi}$ normalizes the integral of the charge dispersion to unity.

The mass yield, $Y_A$, is parameterized as an exponential function of the number of removed nucleons, $A_p-A$:
\begin{equation}\label{eq:YA}
Y_A=SP\exp{\left[-P(A_p-A)\right]}.
\end{equation}
$S$ is the overall scaling factor that accounts for the peripheral nature of the fragmentation reaction and proportional to the sum of the projectile and the target radii:
\begin{equation}
S=S_2(A^{1/3}_p+A^{1/3}_t+S_1).
\end{equation}
with $S_1$ and $S_2$ being fitting parameters. 

The slope of the exponential function in Equation (\ref{eq:YA}), $P$, is taken as a function of the projectile mass, $A_p$, with $P_1$ and $P_2$ as fitting parameters:
\begin{equation}
P=\exp{\left(P_2A_p+P_1\right)}.
\end{equation}

The charge dispersion term, $\exp{\left( -R|Z_{prob}-Z|^{U_{\mathrm{n(p)}}}\right)}$, in Equation (\ref{eq:epax}) is described by three parameters $R$, $Z_{prob}$, and $U_{n(p)}$. These parameters are strongly correlated \cite{sum00}. 

The width parameter, $R$, of the charge distribution is parameterized as a function of the fragment mass, $A$, with $R_1$ and $R_2$ as fitting parameters:
\begin{equation}
R=\exp{\left(R_2A+R_1\right)}.
\end{equation}

To account for the asymmetric nature of the shape of isobaric distributions, the exponents, $U_{\mathrm{n}}$ and $U_{\mathrm{p}}$, for the neutron-rich and proton-rich sides are different. 
\begin{equation}\label{eq:neutron}
U_{\mathrm{n}}=U_{n0} + U_{n1} A
\end{equation}
\begin{equation}\label{eq:proton}
U_{\mathrm{p}}=U_1+U_2 A+U_3 A^2
\end{equation}

The maximum of the isobar distribution, $Z_{prob}$, lies in the valley of stability and it is parameterized as:
\begin{equation}\label{eq:Zprob}
Z_{prob}(A)=Z_{\beta}(A)+\Delta,
\end{equation}
where $Z_{\beta}(A)$ is approximated by a smooth function of the mass number, $A$:
\begin{equation}
Z_{\beta}(A)=\frac{A}{1.98+0.0155A^{2/3}},
\end{equation}
and the $\Delta$ parameter is found to be a linear function of the fragment mass, $A$, for heavy fragments and a quadratic function of $A$ for lower masses:
\begin{equation}\label{eq:delta1}
\Delta=\left\{
\begin{array}{ll}
\Delta_2 A+\Delta_1& \mathrm{if}\, A\geq \Delta_4,\\
\Delta_3 A^2& \mathrm{if}\, A<\Delta_4,\\
\end{array}
\right.
\end{equation}
where $\Delta_1$, $\Delta_2$, $\Delta_3$, and $\Delta_4$ are EPAX parameters. 

The above description from Eq. (\ref{eq:epax}) to (\ref{eq:delta1}) is sufficient to predict the cross sections of fragments located close to the line of stability and far from the projectile nucleus, also referred to as the ``residue corridor.'' For fragments with masses close to the projectile, corrections to the parameters $\Delta$, $R$, and $Y_A$ are introduced, according to the following equations:
\begin{equation}\label{eq:Delta}
\Delta=\Delta\left[ 1+d_1(A/A_p-d_2)^2\right],
\end{equation}
\begin{equation}
R=R\left[ 1+r_1(A/A_p-r_2)^2\right],
\end{equation}
\begin{equation}
Y_A=Y_A\left[ 1+y_1(A/A_p-y_2)^2\right],
\end{equation}
for $(A/A_p-d_2)>0$, $(A/A_p-r_2)>0$, and $(A/A_p-y_2)>0$, respectively.  

A final correction is applied in the case of projectile nuclei far from the line of $\beta$-stability, $Z_{\beta}(A_p)$. In this case, the fragment distributions keep some memory of the $A/Z$ ratio of the projectile nucleus resulting in a correction to the maximum, $Z_{prob}$, of the charge distribution:
\begin{equation}\label{eq:ZprobM}
Z_{prob}(A)=Z_{\beta}(A)+\Delta+\Delta_m,
\end{equation}
where $\Delta_m$ is expressed separately for neutron-rich ($\left(Z_p-Z_{\beta}(A_p)\right)<0$) and proton-rich ($\left(Z_p-Z_{\beta}(A_p)\right) >0$) projectiles:
\begin{equation}\label{eq:DeltaM}
\Delta_m=\left\{
\begin{array}{ll}
(Z_p-Z_{\beta}(A_p))\left[ n_1(A/A_p)^2+n_2(A/A_p)^4\right]&\mathrm{for\, neutron\,rich,}\\
(Z_p-Z_{\beta}(A_p))\exp{\left[ p_1+p_2(A/A_p)\right]}&\mathrm{for\, proton\,rich,}\\
\end{array}
\right.
\end{equation}
where $n_1$, $n_2$ and $p_1$, $p_2$ are fitting parameters. 

The EPAX parameterization altogether contains 24 parameters ($S_1$, $S_2$, $P_1$, $P_2$, $R_1$, $R_2$, $\Delta_1$, $\Delta_2$, $\Delta_3$, $\Delta_4$, $U_{n0}$, $U_{n1}$, $U_1$, $U_2$, $U_3$, $n_1$, $n_2$, $p_1$, $p_2$, $d_1$, $d_2$, $r_1$, $r_2$, $y_1$, and $y_2$), many of which are strongly intercorrelated. The values used are listed in the middle column in Table \ref{table}. 

The present set of data of \nuc{86}{Kr}+\nuc{9}{Be} does not have as extensive mass range as the data from Ref. \cite{web94}. Therefore, Eq. (\ref{eq:delta1}) is reduced to fitting only one mass region with one parameter, $\Delta_3$. Similarly, we do not make corrections to $\Delta$ in Eq. (\ref{eq:Delta}). We also found some improvement if Eq. (\ref{eq:proton}) is mass dependent. (The parameter $U_{n1}$ in that equation was absent in the original EPAX fitting.) All the parameters used in $\mathrm{EPAX_{Kr}}$ are listed in the rightmost column in Table \ref{table}. Note that these are best-fit parameters to our data and cannot not be applied to other reactions or at different energies.

\begin{table*}
\caption{Parameter values for EPAX \cite{sum00} and $\mathrm{EPAX_{Kr}}$.  $\mathrm{EPAX_{Kr}}$ parameters are obtained by fitting the \nuc{86}{Kr}+\nuc{9}{Be} reaction cross-sections.}
\label{table}
\begin{center}
\begin{tabular}{|l|c|c|}
\hline
Parameter&EPAX&$\mathrm{EPAX_{Kr}}$\\
\hline \hline
$S_1$&$-2.38$&0.0\\
$S_2$&0.270&0.431175\\
\hline
$P_1$&$-2.584$&$-2.01932$\\
$P_2$&$-7.5700\times 10^{-3}$&$-1.00263\times 10^{-3}$\\
\hline
$R_1$&0.885&1.4433\\
$R_2$&$-9.8160\times 10^{-3}$&$-2.0546\times 10^{-2}$\\
\hline
$\Delta_1$&-1.087&N/A\\
$\Delta_2$&$3.0470\times 10^{-2}$&N/A\\
$\Delta_3$&$2.1353\times 10^{-4}$&$2.1353\times 10^{-4}$\\
$\Delta_4$&71.35&N/A\\
\hline
$U_{n0}$&1.65&1.7924\\
$U_{n1}$&N/A&$9.819\times 10^{-4}$\\
\hline
$U_1$&1.788&11.284\\
$U_2$&$4.7210\times 10^{-3}$&$-0.2505$\\
$U_3$&$-1.3030\times 10^{-5}$&$1.7676\times 10^{-3}$\\
\hline
$n_1$&0.4&$-0.4$\\
$n_2$&0.6&0.95\\
\hline
$p_1$&$-10.25$&$-10.25$\\
$p_2$&10.1&10.1\\
\hline
$d_1$&$-25.0$&N/A\\
$d_2$&0.80&N/A\\
\hline
$r_1$&20.0&$-1.5$\\
$r_2$&0.82&0.8\\
\hline
$y_1$&200.0&$-10.0$\\
$y_2$&0.90&0.752395\\
\hline
\end{tabular}
\end{center}
\end{table*}
\pagebreak

\bibliographystyle{apsrev}
\bibliography{references}

\begin{thebibliography}{24}
\expandafter\ifx\csname natexlab\endcsname\relax\def\natexlab#1{#1}\fi
\expandafter\ifx\csname bibnamefont\endcsname\relax
  \def\bibnamefont#1{#1}\fi
\expandafter\ifx\csname bibfnamefont\endcsname\relax
  \def\bibfnamefont#1{#1}\fi
\expandafter\ifx\csname citenamefont\endcsname\relax
  \def\citenamefont#1{#1}\fi
\expandafter\ifx\csname url\endcsname\relax
  \def\url#1{\texttt{#1}}\fi
\expandafter\ifx\csname urlprefix\endcsname\relax\def\urlprefix{URL }\fi
\providecommand{\bibinfo}[2]{#2}
\providecommand{\eprint}[2][]{\url{#2}}

\bibitem[{\citenamefont{Tanihata}(1991)}]{tan91}
\bibinfo{author}{\bibfnamefont{I.}~\bibnamefont{Tanihata}},
  \bibinfo{journal}{Nucl. Phys.~A} \textbf{\bibinfo{volume}{{\bf 522}}},
  \bibinfo{pages}{275c} (\bibinfo{year}{1991}).

\bibitem[{\citenamefont{Tanihata et~al.}(1992)\citenamefont{Tanihata, Hirata,
  Kobayashi, Shimoura, Sugimoto, and Toki}}]{tan92}
\bibinfo{author}{\bibfnamefont{I.}~\bibnamefont{Tanihata}},
  \bibinfo{author}{\bibfnamefont{D.}~\bibnamefont{Hirata}},
  \bibinfo{author}{\bibfnamefont{T.}~\bibnamefont{Kobayashi}},
  \bibinfo{author}{\bibfnamefont{S.}~\bibnamefont{Shimoura}},
  \bibinfo{author}{\bibfnamefont{K.}~\bibnamefont{Sugimoto}}, \bibnamefont{and}
  \bibinfo{author}{\bibfnamefont{H.}~\bibnamefont{Toki}},
  \bibinfo{journal}{Phys. Lett.~B} \textbf{\bibinfo{volume}{{\bf 289}}},
  \bibinfo{pages}{261} (\bibinfo{year}{1992}).

\bibitem[{\citenamefont{Fukunishi et~al.}(1993)\citenamefont{Fukunishi, Otsuka,
  and Tanihata}}]{fuk93}
\bibinfo{author}{\bibfnamefont{N.}~\bibnamefont{Fukunishi}},
  \bibinfo{author}{\bibfnamefont{T.}~\bibnamefont{Otsuka}}, \bibnamefont{and}
  \bibinfo{author}{\bibfnamefont{I.}~\bibnamefont{Tanihata}},
  \bibinfo{journal}{Phys. Rev.~C} \textbf{\bibinfo{volume}{{\bf 48}}},
  \bibinfo{pages}{1648} (\bibinfo{year}{1993}).

\bibitem[{\citenamefont{Motobayashi et~al.}(1995)\citenamefont{Motobayashi,
  Ikeda, Ando, Ieki, Inoue, Iwasa, Kikuchi, Kurokawa, Moryia, Ogawa
  et~al.}}]{mot95}
\bibinfo{author}{\bibfnamefont{T.}~\bibnamefont{Motobayashi}},
  \bibinfo{author}{\bibfnamefont{Y.}~\bibnamefont{Ikeda}},
  \bibinfo{author}{\bibfnamefont{Y.}~\bibnamefont{Ando}},
  \bibinfo{author}{\bibfnamefont{K.}~\bibnamefont{Ieki}},
  \bibinfo{author}{\bibfnamefont{M.}~\bibnamefont{Inoue}},
  \bibinfo{author}{\bibfnamefont{N.}~\bibnamefont{Iwasa}},
  \bibinfo{author}{\bibfnamefont{T.}~\bibnamefont{Kikuchi}},
  \bibinfo{author}{\bibfnamefont{M.}~\bibnamefont{Kurokawa}},
  \bibinfo{author}{\bibfnamefont{W.}~\bibnamefont{Moryia}},
  \bibinfo{author}{\bibfnamefont{S.}~\bibnamefont{Ogawa}},
  \bibnamefont{et~al.}, \bibinfo{journal}{Phys. Lett.~B}
  \textbf{\bibinfo{volume}{{\bf 346}}}, \bibinfo{pages}{9}
  (\bibinfo{year}{1995}).

\bibitem[{\citenamefont{Yano}(1989)}]{yan89}
\bibinfo{author}{\bibfnamefont{Y.}~\bibnamefont{Yano}}, in
  \emph{\bibinfo{booktitle}{Proceedings 12th Int. Conf. on Cyclotrons and their
  applications}}, edited by
  \bibinfo{editor}{\bibfnamefont{B.}~\bibnamefont{Martin}} \bibnamefont{and}
  \bibinfo{editor}{\bibfnamefont{K.}~\bibnamefont{Ziegler}}
  (\bibinfo{publisher}{Word Scientific}, \bibinfo{year}{1989}).

\bibitem[{\citenamefont{Kubo et~al.}(1992)\citenamefont{Kubo, Ishihara, Inabe,
  Kumagai, Tanihata, and Yoshida}}]{kub92}
\bibinfo{author}{\bibfnamefont{T.}~\bibnamefont{Kubo}},
  \bibinfo{author}{\bibfnamefont{M.}~\bibnamefont{Ishihara}},
  \bibinfo{author}{\bibfnamefont{N.}~\bibnamefont{Inabe}},
  \bibinfo{author}{\bibfnamefont{H.}~\bibnamefont{Kumagai}},
  \bibinfo{author}{\bibfnamefont{I.}~\bibnamefont{Tanihata}}, \bibnamefont{and}
  \bibinfo{author}{\bibfnamefont{K.}~\bibnamefont{Yoshida}},
  \bibinfo{journal}{Nucl. Instrum. Methods Phys. Res., Sect.~B}
  \textbf{\bibinfo{volume}{{\bf 70}}}, \bibinfo{pages}{309}
  (\bibinfo{year}{1992}).

\bibitem[{\citenamefont{Mocko}(2006)}]{moc06}
\bibinfo{author}{\bibfnamefont{M.}~\bibnamefont{Mocko}}, Ph.D. thesis,
  \bibinfo{school}{Michigan State University} (\bibinfo{year}{2006}).

\bibitem[{\citenamefont{Scheidenberg et~al.}(1998)\citenamefont{Scheidenberg,
  St\"ohlker, Meyerhof, Geissel, Mokler, and Blank}}]{sch98}
\bibinfo{author}{\bibfnamefont{C.}~\bibnamefont{Scheidenberg}},
  \bibinfo{author}{\bibfnamefont{T.}~\bibnamefont{St\"ohlker}},
  \bibinfo{author}{\bibfnamefont{W.~E.} \bibnamefont{Meyerhof}},
  \bibinfo{author}{\bibfnamefont{H.}~\bibnamefont{Geissel}},
  \bibinfo{author}{\bibfnamefont{P.~H.} \bibnamefont{Mokler}},
  \bibnamefont{and} \bibinfo{author}{\bibfnamefont{B.}~\bibnamefont{Blank}},
  \bibinfo{journal}{Nucl. Instrum. Methods Phys. Res., Sect.~B}
  \textbf{\bibinfo{volume}{{\bf 142}}}, \bibinfo{pages}{441}
  (\bibinfo{year}{1998}).

\bibitem[{\citenamefont{Bazin et~al.}(2002)\citenamefont{Bazin, Tarasov,
  Lewitowicz, and Sorlin}}]{baz02}
\bibinfo{author}{\bibfnamefont{D.}~\bibnamefont{Bazin}},
  \bibinfo{author}{\bibfnamefont{O.}~\bibnamefont{Tarasov}},
  \bibinfo{author}{\bibfnamefont{M.}~\bibnamefont{Lewitowicz}},
  \bibnamefont{and} \bibinfo{author}{\bibfnamefont{O.}~\bibnamefont{Sorlin}},
  \bibinfo{journal}{Nucl. Instrum. Methods Phys. Res., Sect.~A}
  \textbf{\bibinfo{volume}{{\bf 482}}}, \bibinfo{pages}{307}
  (\bibinfo{year}{2002}), \urlprefix\url{http://www.nscl.msu.edu/lise}.

\bibitem[{\citenamefont{Bazin et~al.}(1990)\citenamefont{Bazin, Guerreau, Anne,
  Guillemaud-Mueller, Mueller, and Saint-Laurent}}]{baz90}
\bibinfo{author}{\bibfnamefont{D.}~\bibnamefont{Bazin}},
  \bibinfo{author}{\bibfnamefont{D.}~\bibnamefont{Guerreau}},
  \bibinfo{author}{\bibfnamefont{R.}~\bibnamefont{Anne}},
  \bibinfo{author}{\bibfnamefont{D.}~\bibnamefont{Guillemaud-Mueller}},
  \bibinfo{author}{\bibfnamefont{A.~C.} \bibnamefont{Mueller}},
  \bibnamefont{and} \bibinfo{author}{\bibfnamefont{M.~G.}
  \bibnamefont{Saint-Laurent}}, \bibinfo{journal}{Nucl. Phys.~A}
  \textbf{\bibinfo{volume}{{\bf 515}}}, \bibinfo{pages}{349}
  (\bibinfo{year}{1990}).

\bibitem[{\citenamefont{Mocko et~al.}(2006)\citenamefont{Mocko, Tsang,
  Andronenko, Andronenko, Delaunay, Famiano, Ginter, Henzl, Henzlov\'a, Hua
  et~al.}}]{moc06a}
\bibinfo{author}{\bibfnamefont{M.}~\bibnamefont{Mocko}},
  \bibinfo{author}{\bibfnamefont{M.~B.} \bibnamefont{Tsang}},
  \bibinfo{author}{\bibfnamefont{L.}~\bibnamefont{Andronenko}},
  \bibinfo{author}{\bibfnamefont{M.}~\bibnamefont{Andronenko}},
  \bibinfo{author}{\bibfnamefont{F.}~\bibnamefont{Delaunay}},
  \bibinfo{author}{\bibfnamefont{M.~A.} \bibnamefont{Famiano}},
  \bibinfo{author}{\bibfnamefont{T.}~\bibnamefont{Ginter}},
  \bibinfo{author}{\bibfnamefont{V.}~\bibnamefont{Henzl}},
  \bibinfo{author}{\bibfnamefont{D.}~\bibnamefont{Henzlov\'a}},
  \bibinfo{author}{\bibfnamefont{H.}~\bibnamefont{Hua}}, \bibnamefont{et~al.},
  \bibinfo{journal}{Phys. Rev.~C} \textbf{\bibinfo{volume}{{\bf 74}}},
  \bibinfo{pages}{054612} (\bibinfo{year}{2006}).

\bibitem[{\citenamefont{Iwasa et~al.}(1997)\citenamefont{Iwasa, Geissel,
  M\"unzenberg, Scheidenberger, Schwab, and Wollnik}}]{iwa97}
\bibinfo{author}{\bibfnamefont{N.}~\bibnamefont{Iwasa}},
  \bibinfo{author}{\bibfnamefont{H.}~\bibnamefont{Geissel}},
  \bibinfo{author}{\bibfnamefont{G.}~\bibnamefont{M\"unzenberg}},
  \bibinfo{author}{\bibfnamefont{C.}~\bibnamefont{Scheidenberger}},
  \bibinfo{author}{\bibfnamefont{T.}~\bibnamefont{Schwab}}, \bibnamefont{and}
  \bibinfo{author}{\bibfnamefont{H.}~\bibnamefont{Wollnik}},
  \bibinfo{journal}{Nucl. Instrum. Methods Phys. Res., Sect.~B}
  \textbf{\bibinfo{volume}{{\bf 126}}}, \bibinfo{pages}{284}
  (\bibinfo{year}{1997}).

\bibitem[{\citenamefont{Bibber et~al.}(1979)\citenamefont{Bibber, Hendrie,
  Scott, Weiman, Schroeder, Geaga, Cessin, Treuhaft, Grossiord, Rasmussen
  et~al.}}]{bib79}
\bibinfo{author}{\bibfnamefont{K.~V.} \bibnamefont{Bibber}},
  \bibinfo{author}{\bibfnamefont{D.~L.} \bibnamefont{Hendrie}},
  \bibinfo{author}{\bibfnamefont{D.~K.} \bibnamefont{Scott}},
  \bibinfo{author}{\bibfnamefont{H.~H.} \bibnamefont{Weiman}},
  \bibinfo{author}{\bibfnamefont{L.~S.} \bibnamefont{Schroeder}},
  \bibinfo{author}{\bibfnamefont{J.~V.} \bibnamefont{Geaga}},
  \bibinfo{author}{\bibfnamefont{S.~A.} \bibnamefont{Cessin}},
  \bibinfo{author}{\bibfnamefont{R.}~\bibnamefont{Treuhaft}},
  \bibinfo{author}{\bibfnamefont{Y.~J.} \bibnamefont{Grossiord}},
  \bibinfo{author}{\bibfnamefont{J.~O.} \bibnamefont{Rasmussen}},
  \bibnamefont{et~al.}, \bibinfo{journal}{Phys. Rev. Lett.}
  \textbf{\bibinfo{volume}{{\bf 43}}}, \bibinfo{pages}{840}
  (\bibinfo{year}{1979}).

\bibitem[{\citenamefont{Goldhaber}(1974)}]{gold74}
\bibinfo{author}{\bibfnamefont{A.~S.} \bibnamefont{Goldhaber}},
  \bibinfo{journal}{Phys. Lett.~B} \textbf{\bibinfo{volume}{{\bf 53}}},
  \bibinfo{pages}{306} (\bibinfo{year}{1974}).

\bibitem[{\citenamefont{Dayras et~al.}(1986)\citenamefont{Dayras, Pagano,
  Barrette, Berthier, Rizzo, Chavez, Cisse, Legrain, Mermaz, Pollacco
  et~al.}}]{day86}
\bibinfo{author}{\bibfnamefont{R.}~\bibnamefont{Dayras}},
  \bibinfo{author}{\bibfnamefont{A.}~\bibnamefont{Pagano}},
  \bibinfo{author}{\bibfnamefont{J.}~\bibnamefont{Barrette}},
  \bibinfo{author}{\bibfnamefont{B.}~\bibnamefont{Berthier}},
  \bibinfo{author}{\bibfnamefont{D.~M. D.~C.} \bibnamefont{Rizzo}},
  \bibinfo{author}{\bibfnamefont{E.}~\bibnamefont{Chavez}},
  \bibinfo{author}{\bibfnamefont{O.}~\bibnamefont{Cisse}},
  \bibinfo{author}{\bibfnamefont{R.}~\bibnamefont{Legrain}},
  \bibinfo{author}{\bibfnamefont{M.~C.} \bibnamefont{Mermaz}},
  \bibinfo{author}{\bibfnamefont{E.~C.} \bibnamefont{Pollacco}},
  \bibnamefont{et~al.}, \bibinfo{journal}{Nucl. Phys.~A}
  \textbf{\bibinfo{volume}{{\bf 460}}}, \bibinfo{pages}{299}
  (\bibinfo{year}{1986}).

\bibitem[{\citenamefont{Notani et~al.}()\citenamefont{Notani, Sakurai, Aoi,
  Iwasaki, Fukuda, Liu, Yoneda, Ogawa, Teranishi, Nakamura et~al.}}]{not07}
\bibinfo{author}{\bibfnamefont{M.}~\bibnamefont{Notani}},
  \bibinfo{author}{\bibfnamefont{H.}~\bibnamefont{Sakurai}},
  \bibinfo{author}{\bibfnamefont{N.}~\bibnamefont{Aoi}},
  \bibinfo{author}{\bibfnamefont{H.}~\bibnamefont{Iwasaki}},
  \bibinfo{author}{\bibfnamefont{N.}~\bibnamefont{Fukuda}},
  \bibinfo{author}{\bibfnamefont{Z.}~\bibnamefont{Liu}},
  \bibinfo{author}{\bibfnamefont{K.}~\bibnamefont{Yoneda}},
  \bibinfo{author}{\bibfnamefont{H.}~\bibnamefont{Ogawa}},
  \bibinfo{author}{\bibfnamefont{T.}~\bibnamefont{Teranishi}},
  \bibinfo{author}{\bibfnamefont{T.}~\bibnamefont{Nakamura}},
  \bibnamefont{et~al.}, \bibinfo{note}{nucl-ex/0702050v1}.

\bibitem[{\citenamefont{Kox et~al.}(1984)\citenamefont{Kox, Gamp, Cherkaoui,
  Cole, Longequeue, Menet, Perrin, and Viano}}]{kox84}
\bibinfo{author}{\bibfnamefont{S.}~\bibnamefont{Kox}},
  \bibinfo{author}{\bibfnamefont{A.}~\bibnamefont{Gamp}},
  \bibinfo{author}{\bibfnamefont{P.}~\bibnamefont{Cherkaoui}},
  \bibinfo{author}{\bibfnamefont{A.~J.} \bibnamefont{Cole}},
  \bibinfo{author}{\bibfnamefont{N.}~\bibnamefont{Longequeue}},
  \bibinfo{author}{\bibfnamefont{J.}~\bibnamefont{Menet}},
  \bibinfo{author}{\bibfnamefont{C.}~\bibnamefont{Perrin}}, \bibnamefont{and}
  \bibinfo{author}{\bibfnamefont{J.~B.} \bibnamefont{Viano}},
  \bibinfo{journal}{Nucl. Phys.~A} \textbf{\bibinfo{volume}{{\bf 420}}},
  \bibinfo{pages}{162} (\bibinfo{year}{1984}).

\bibitem[{\citenamefont{Sakurai et~al.}(1999)\citenamefont{Sakurai, Lukyanov,
  Notani, Aoi, Beaumel, Fukuda, Hirai, Ideguchi, Imai, Ishihara
  et~al.}}]{sak99}
\bibinfo{author}{\bibfnamefont{H.}~\bibnamefont{Sakurai}},
  \bibinfo{author}{\bibfnamefont{S.~M.} \bibnamefont{Lukyanov}},
  \bibinfo{author}{\bibfnamefont{M.}~\bibnamefont{Notani}},
  \bibinfo{author}{\bibfnamefont{N.}~\bibnamefont{Aoi}},
  \bibinfo{author}{\bibfnamefont{D.}~\bibnamefont{Beaumel}},
  \bibinfo{author}{\bibfnamefont{N.}~\bibnamefont{Fukuda}},
  \bibinfo{author}{\bibfnamefont{M.}~\bibnamefont{Hirai}},
  \bibinfo{author}{\bibfnamefont{E.}~\bibnamefont{Ideguchi}},
  \bibinfo{author}{\bibfnamefont{N.}~\bibnamefont{Imai}},
  \bibinfo{author}{\bibfnamefont{M.}~\bibnamefont{Ishihara}},
  \bibnamefont{et~al.}, \bibinfo{journal}{Phys. Lett.~B}
  \textbf{\bibinfo{volume}{{\bf 448}}}, \bibinfo{pages}{180}
  (\bibinfo{year}{1999}).

\bibitem[{\citenamefont{S\"ummerer}(2003)}]{sum03}
\bibinfo{author}{\bibfnamefont{K.}~\bibnamefont{S\"ummerer}},
  \bibinfo{journal}{Nucl. Instrum. Methods Phys. Res., Sect.~B}
  \textbf{\bibinfo{volume}{{\bf 204}}}, \bibinfo{pages}{278}
  (\bibinfo{year}{2003}).

\bibitem[{\citenamefont{Weber et~al.}(1994)\citenamefont{Weber, Donzaud,
  Dufour, Geissel, Grewe, Guillemaud-Mueller, Keller, Lewitowicz, Magel,
  Mueller et~al.}}]{web94}
\bibinfo{author}{\bibfnamefont{M.}~\bibnamefont{Weber}},
  \bibinfo{author}{\bibfnamefont{C.}~\bibnamefont{Donzaud}},
  \bibinfo{author}{\bibfnamefont{J.~P.} \bibnamefont{Dufour}},
  \bibinfo{author}{\bibfnamefont{H.}~\bibnamefont{Geissel}},
  \bibinfo{author}{\bibfnamefont{A.}~\bibnamefont{Grewe}},
  \bibinfo{author}{\bibfnamefont{D.}~\bibnamefont{Guillemaud-Mueller}},
  \bibinfo{author}{\bibfnamefont{H.}~\bibnamefont{Keller}},
  \bibinfo{author}{\bibfnamefont{M.}~\bibnamefont{Lewitowicz}},
  \bibinfo{author}{\bibfnamefont{A.}~\bibnamefont{Magel}},
  \bibinfo{author}{\bibfnamefont{A.~C.} \bibnamefont{Mueller}},
  \bibnamefont{et~al.}, \bibinfo{journal}{Nucl. Phys.~A}
  \textbf{\bibinfo{volume}{{\bf 578}}}, \bibinfo{pages}{659}
  (\bibinfo{year}{1994}).

\bibitem[{\citenamefont{S\"ummerer and Blank}(2000)}]{sum00}
\bibinfo{author}{\bibfnamefont{K.}~\bibnamefont{S\"ummerer}} \bibnamefont{and}
  \bibinfo{author}{\bibfnamefont{B.}~\bibnamefont{Blank}},
  \bibinfo{journal}{Phys. Rev.~C} \textbf{\bibinfo{volume}{{\bf 61}}},
  \bibinfo{pages}{034607} (\bibinfo{year}{2000}).

\bibitem[{\citenamefont{Friedman and Tsang}(2003)}]{fri03}
\bibinfo{author}{\bibfnamefont{W.~A.} \bibnamefont{Friedman}} \bibnamefont{and}
  \bibinfo{author}{\bibfnamefont{M.~B.} \bibnamefont{Tsang}},
  \bibinfo{journal}{Phys. Rev.~C} \textbf{\bibinfo{volume}{{\bf 67}}},
  \bibinfo{pages}{051601} (\bibinfo{year}{2003}).

\bibitem[{\citenamefont{S\"ummerer}()}]{sum06}
\bibinfo{author}{\bibfnamefont{K.}~\bibnamefont{S\"ummerer}},
  \bibinfo{howpublished}{private communications}.

\bibitem[{\citenamefont{Sun}()}]{sun07}
\bibinfo{author}{\bibfnamefont{Z.~Y.} \bibnamefont{Sun}},
  \bibinfo{howpublished}{private communications}.

\end{thebibliography}

\end{document}